\documentclass[12pt]{article}
\pdfoutput=1 
\usepackage{hyperref}

\usepackage{amsmath}
\usepackage{amssymb}
\usepackage{graphicx}
\usepackage{enumerate}
\usepackage{slashed, bm}
\usepackage{xcolor}
\usepackage[final]{showlabels}
\definecolor{dark}{rgb}{0.10,0.2,0.3}
\definecolor{magenta}{rgb}{0.7,0.1,0.3}
\definecolor{purpure}{rgb}{0.5,0.15,0.3}
\usepackage[font=small,format=plain,labelfont=bf,up,textfont=it,up]{caption}
\usepackage{hyperref, cite}
\hypersetup{colorlinks,
citecolor=blue,
filecolor=blue,
linkcolor=magenta,
urlcolor=purpure,hyperfootnotes=true,pdftex}

\let\a = \alpha    \let\b =\beta     \let\g = \gamma   \let\d = \delta        
            \let\k =\kappa      
                    \let\r = \rho            

\newcommand{\pd}{\partial}
\newcommand{\nn}{\nonumber}

\newcommand{\beq}{\begin{equation}}
\newcommand{\eeq}{\end{equation}}
\newcommand{\bea}{\begin{eqnarray}}
\newcommand{\eea}{\end{eqnarray}}
\newcommand{\llangle}{\left\langle}
\newcommand{\rrangle}{\right\rangle}

\newcommand{\muu}{\mu_{1}}
\newcommand{\nuu}{\nu_{1}}

\newcommand{\pu}{p_1}
\newcommand{\ku}{k_1}
\newcommand{\xu}{x_{1}}

\newcommand{\mud}{\mu_{2}}
\newcommand{\nud}{\nu_{2}}

\newcommand{\pdd}{p_2}
\newcommand{\kd}{k_2}
\newcommand{\xd}{x_{2}}

\newcommand{\mut}{\mu_{3}}
\newcommand{\nut}{\nu_{3}}

\newcommand{\pt}{p_3}
\newcommand{\kt}{k_3}
\newcommand{\xt}{x_{3}}

\newcommand{\muq}{\mu_{4}}
\newcommand{\nuq}{\nu_{4}}

\newcommand{\pq}{p_4}

\newcommand{\xq}{x_{4}}

\newcommand{\wolfram}{\emph{Mathematica} }
\newcommand{\patel}{\emph{Package-X} }
\newcommand{\denner}{\emph{CollierLink} }

\title{
\vspace{-4.0cm}
\begin{flushright}
\end{flushright}
\vspace{1.5cm}
\boldmath \large \bf The four-point correlation function of the energy-momentum tensor \\ in the free conformal field theory of a scalar field}

\author{
Mirko~Serino \\
Department of Physics, Ben Gurion University of the Negev \\ Beer Sheva 8410501, Israel\\
\emph{Email}: mirkos.serino@gmail.com
}

\begin{document}

\maketitle
\flushbottom
\abstract{
We present an explicit momentum space computation of the four-point function of the energy-momentum tensor in 4 spacetime dimensions
for the free and conformally invariant theory of a scalar field. The result is obtained by explicit evaluation of the Feynman diagrams by tensor reduction.
We work by embedding the scalar field theory in a gravitational background consistently with conformal invariance in order to derive 
all the terms the correlator consists of and all the Ward identities implied by the requirements of general covariance and anomalous Weyl symmetry.
We test all these identities numerically in several kinematic configurations.
\wolfram notebooks detailing the step-by-step computation are made publicly available through a GitHub repository\footnote{\url{https://github.com/mirkos86/4-EMT-correlation-function-in-a-4d-CFT}}.
To the best of our knowledge, this is the first explicit result for the four-point correlation function of the energy-momentum tensor in a conformal 
and non supersymmetric field theory which is readily numerically evaluable in any kinematic configuration.
}

\vspace{3cm}
\begin{flushright}
In loving memory of Giuseppe Serino\\ 
*\, April 1st 1956 \,\,\, \textdagger \,December 4th 2018 \\
\end{flushright}

\newpage

\tableofcontents

\section{Introduction}\label{intro}

The energy-momentum tensor is the most universal operator for conformal field theories (CFTs).
Therefore, its correlation functions are natural objects to study in any CFT.

It has been known for a long time that conformal invariance completely fixes the two-point correlation function of the energy-momentum tensor, whereas
the three-point function in 4d is fixed up to three scalar coefficients, which can be inferred by computing the correlator for the field theories of a scalar, a fermion 
and an abelian gauge field. 
This was first dealt with in coordinate space for three-point correlators of operators up to spin 2 in~\cite{Osborn:1993cr,Erdmenger:1996yc}, 
by solving all the constraints implied by the full conformal group. 
An analogous program in momentum space, based on the solution of all the conformal group Ward identities to constrain the 3 point functions of scalars, 
vector currents and energy-momentum tensors modulo a few constants was carried out much more recently~\cite{Coriano:2013jba,Bzowski:2013sza,Bzowski:2017poo}
and found to be perfectly consistent with the other approach, although mathematically very different. 
A different approach in momentum space to the three-point function of the energy-momentum tensor has been pursued in~\cite{Coriano:2012wp}, 
where an explicit diagrammatic computation was performed (and published in a partially on-shell kinematics). 
The consistency of the two momentum space approaches was later verified by~\cite{Coriano:2018bsy}.

Much more involved is the case for four-point functions, because the conformal group does not uniquely constrain them.
For the case of scalars, Polyakov first showed that the four-point correlator is fixed up to a function of the coordinates cross ratios~\cite{Polyakov:1970xd},
\bea
&&
\langle  \mathcal O(\xu)\,\mathcal O(\xd)\,\mathcal O(\xt)\,\mathcal O(\xq)\, \rangle =
\frac{f_I(u,v)}{(\xu-\xd)^{2\Delta}\, (\xt-\xq)^{2\Delta}} \, , 
\nn \\
&&
u = \frac{(\xu-\xd)^2\,(\xt-\xq)^2}{(\xu-\xt)^2\,(\xd-\xq)^2} \, , v = \frac{(\xu-\xq)^2\,(\xd-\xt)^2}{(\xu-\xt)^2\,(\xd-\xq)^2} \, .
\eea
As for operators with spin, though much is already known in position space~\cite{Costa:2011mg}, very little is known about four-point correlation functions in momentum space
and the connection between the two is not trivial, as discussed thoroughly in~\cite{Coriano:2012wp}. 
One very powerful and popular approach to CFT dynamics is the so called conformal bootstrap~\cite{Ferrara:1973yt,polyakov:1974gs,Simmons-Duffin:2016gjk},
which allows to constrain four-point functions through consistency constraints connecting different conformal blocks representations
of the same four-point function. So far, it has been successfully applied mainly to the case of scalar operators. 
One exception is a very recently proposed approach to conformal blocks for any spin~\cite{Fortin:2016lmf,Fortin:2019fvx}, 
which was applied to two, three and four-point fucntions~\cite{Fortin:2019xyr,Fortin:2019pep,Fortin:2019gck},
though not to to the four-point function of the energy-momentum tensor due to its sheer complexity.

The number of unrestricted degrees of freedom for the four-point function of the energy-momentum tensor in general dimensions
was already determined in~\cite{Dymarsky:2013wla}, where the number of effectively independent constraints provided by the Ward identities
stemming from the requirements of both special conformal invariance and energy-momentum conservation was nailed down. 
There have been works - all of them in the context of the AdS/CFT correspondence - which have investigated the four-point function of the energy-momentum tensor. 
The first one has explored the structure of some contributions to the four-point function through an OPE analysis~\cite{Chalmers:2000vq} .
Later, the most general structure in coordinate space of the correlator was pinned down in~\cite{Didenko:2012tv,Didenko:2013bj}. 
By different means, analogous structures were conjectured in~\cite{Sleight:2016dba} and later proven to be correct 
in~\cite{Bonezzi:2017vha}.\footnote{We are grateful to Evgeny Skvortsov and Johanna Erdmenger for pointing out these references.}
So far, however, no explicit result has been presented in momentum space, 
except for the case of superconformal $\mathcal N = 4$~\cite{Korchemsky:2015ssa}, let alone any which lends itself to numerical evaluation. 

With these recent advances in mind, in the present paper we present an explicit perturbative computation of the four-point function of the energy-momentum tensor
in the simplest possible conformal field theory in 4d which admits a Lagrangian realization and thus a perturbative approach: a free scalar field.
It is paramount to notice that the fact that the theory is free does not just make calculations simpler;
it is such that the 1 loop results for the four-point function (or lower and higher point functions as well), 
which can be computed using perturbation theory, are also exact, because no higher order loops are admitted. 
In this way, the perturbative calculation produces a result equivalent to what would be produced by non perturbative methods,
though unfortunately way less compact because of redundancies.

We provide our results in a set of ancillary \wolfram files stored in the public GitHub repository quoted in the abstract. 
In these files, the reader is guided step by step through the computation of the two, three and four-point correlators of the energy-momentum tensor
for the theory at hand, as well as through the tests of the Ward identities stemming from the requirements of energy-momentum conservation and anomalous Weyl symmetry.
The \wolfram notebooks code features plenty of comments which make them readily usable and (hopefully) understandable. 
We exploit the recently developed \patel~\cite{Patel:2015tea,Patel:2016fam} together with its \denner extension for fast numerical evaluation 
of tensor integrals through the COLLIER library~\cite{Denner:2002ii,Denner:2005nn,Denner:2010tr,Denner:2016kdg}.

We did not attempt to track down the few~\cite{Dymarsky:2013wla} independent tensor structures our correlator should ultimately consist of, by itself a very demanding task 
given the sheer dimensions of the four-point function. We leave an attempt in this direction for possible future work.

The main reason why we undertook this calculation, beside the fact that we could do it, is the hope that our result could serve
as a benchmark check (most probably numerical, though the result itself is fully analytical) for computations
carried out with non perturbative methods, e.g. the conformal bootstrap for fields with spin. \\

This paper is organized as follows. 
In section \ref{setup} we describe the setup of our calculation and the diagrammatic expansion of all the energy-momentum tensor correlators through four point,
in order to make our discussion as self-contained as possible, beside providing a detailed explanation of the structure of the \wolfram files.
In section \ref{ward} we derive all the Ward identities stemming from the requirements of general covariance (named transverse Ward identities hereafter)
and Weyl invariance (named trace Ward identities hereafter), the latter of which feature a well known trace anomaly.  
In section \ref{counterterm} we describe the computation of the 1 loop countertems and the anomaly functional for the energy-momentum tensor correlators through four 
point and how we can check them against each other independently of the Feynman diagram computations; the countertems are then used 
for a preliminary check of the diagrammatic expansion, by comparing them to the UV pole of the correlators. 
In section \ref{mathematica} we illustrate in detail the \wolfram implementation of our calculation and the procedure followed to analytically test the Ward identities
for the two and three-point correlators and to test them numerically for the four-point correlator.
Indeed, the last task is too massive to be dealt with analytically. In this section we also give plenty of 
detail on the implementation of all our computations in \wolfram through \patel and \denner.
Section \ref{ending} presents our conclusions and perspectives for further work.

\section{Setting up the computation}\label{setup}

This introductory section closely follows Chapter 2 of~\cite{Serino:2014ysa}. 
For details about the general relation between conformal invariance in flat space and Weyl invariance in curved space see~\cite{Iorio:1996ad};
here we assume that Weyl invariance in curved space is interchangeable with conformal invariance in flat space, 
as it is actually the case for the theory we are dealing with.

The standard definition of the energy-momentum tensor (EMT in the following)  in a classical field theory described by an action $\mathcal S$ is 
given in terms of a functional derivative w.r.t. the metric tensor $g_{\mu\nu}(z)$ once the theory has been 
embedded in curved space, i.e.
\beq \label{emt}
T^{\mu\nu}(z) = -\frac{2}{\sqrt{g_z}}\frac{\delta \mathcal{S}}{\delta g_{\mu\nu}(z)} = 
g^{\mu\alpha}(z)\,g^{\nu\beta}(z)\,\frac{2}{\sqrt{g_z}}\,\frac{\delta\mathcal S}{\delta g^{\alpha\beta}(z)} \, ,
\eeq
where $\textrm{det}\, g_{\mu\nu}(z)\equiv g_z$.

We introduce the generating functional of the theory in Euclidean conventions, which we call $\mathcal W$,
\beq
\mathcal W = \frac{1}{\mathcal{N}} \, \int \, \mathcal D\Phi \, e^{- \mathcal S}\, ,
\label{genfunc}
\eeq
where $\mathcal{N}$ is a normalization constant and $\Phi$ generally indicates the quantum fields of the theory.
Given (\ref{emt}), the vacuum expectation value (vev) of the EMT is
\beq \label{vevemt}
\llangle T^{\mu\nu}(z) \rrangle_s = \frac{2}{\sqrt{g_z}}\frac{\delta\, \mathcal W}{\delta\, g_{\mu\nu}(z)}\,,
\eeq
with the subscript $s$ (which stands for "source") meaning that the background fields are not switched off.
Dependence on coordinates will be occasionally dropped if it is not strictly necessary to the understanding of formulas.

In a conformal field theory the trace of the EMT vanishes at the classical level upon using the equations of motion, ${T^\mu}_\mu = 0$.
As for the vev of the EMT, this relation is modified by the well known trace anomaly~\cite{Duff:1977ay,Birrell:1982ix}
\beq \label{traceanomaly}
g_{\mu\nu}\llangle T^{\mu\nu} \rrangle_s  \equiv \mathcal{A}[g]  =
\sum_{I=f,s,V}n_I \, \bigg[ \beta_a(I)\, F + \beta_b(I)\, G + \beta_c(I)\,\Box R + \beta_d(I)\, R^2  \bigg] \, ,
\eeq
where $g$ is a short-hand notation for the background metric. 
The coefficients $\kappa$, $\beta_a$, $\beta_b$, $\beta_c$ and $\beta_d$ depend on the field content of the Lagrangian theory
and we have a multiplicity factor $n_I$ for each particle species. 
Eq. (\ref{traceanomaly}) is a reorganization in terms of the squared Weyl tensor and the Euler characteristic of 4d spacetime (see Appendix \ref{Geometrical})
of the most general linear combination of the squares of the Riemann tensor and its contractions. 
The coefficient of $R^2$ must actually vanish identically
\beq
\beta_d \equiv 0 \, , 
\eeq
since it does not satisfy the Wess-Zumino consistency condition for conformal anomalies~\cite{Bonora:1983ff,Antoniadis:1992xu}. 
In addition, the value of $\beta_c$ is regularization dependent, corresponding to 
the fact that it can be changed by the addition of an arbitrary local term in the effective action proportional to the integral of $R^2$~\cite{Duff:1977ay}.
In particular, the values for the anomaly coefficients which we will use here for one single scalar field, i.e.
\beq
\beta_a = \frac{3}{5760\,\pi^2}\, , \quad \beta_b =-  \frac{1}{5760\,\pi^2}
\eeq
for which one finds the constraint \cite{Duff:1977ay,Birrell:1982ix}
\beq
\beta_c = -\frac{2}{3}\,\beta_a \, .
\eeq
The trace anomaly functional only depends on the metric tensor, $\mathcal A \equiv \mathcal{A}[g]$,
\beq
g_{\mu\nu}\llangle T^{\mu\nu} \rrangle_s  =\beta_a\, \left(F - \frac{2}{3} \,\Box R\right) + \beta_b\, G
\label{traceanomalyfinal}
\eeq

The conformally invariant action for a scalar field coupled to gravity in $4$ dimensions is given by 
\beq
\mathcal{S} =
\frac{1}{2} \, \int d^4 x \, \sqrt{g}\, \bigg[g^{\mu\nu}\,\nabla_\mu\phi\,\nabla_\nu\phi - \chi \, R \,\phi^2 \bigg]\, .
\label{scalarAction}
\eeq
Here $\chi$ is the parameter corresponding to the term of improvement obtained by coupling $\phi^2$ to the scalar curvature $R$.
In general dimensions $d$, $\chi = 1/4\, (d-2)/(d-1)$ gives a CFT, so that $\chi = 1/6$ gives a classically conformal invariant theory in $d=4$, 
whose EMT is
\beq
T^{\mu\nu}_{S} =
\nabla^\mu \phi \, \nabla^\nu\phi - \frac{1}{2} \, g^{\mu\nu}\,g^{\alpha\beta}\,\nabla_\alpha \phi \, \nabla_\beta \phi
+ \chi \bigg[g^{\mu\nu} \Box - \nabla^\mu\,\nabla^\nu + \frac{1}{2}\,g^{\mu\nu}\,R - R^{\mu\nu} \bigg]\, \phi^2 \, .
\label{ScalarEMT}
\eeq
The explicit expressions for the vertices involving one or more metric tensors, which can be computed by functional differentiating the action, 
have been already given in~\cite{Coriano:2012hd} and are also here collected in Appendix \ref{Vertices} for completeness, 
beside being explicitly computed in the attached \wolfram file \emph{functional\_derivatives.nb}.
For further details about our conventions on covariant derivatives, Christoffel symbols and the Riemann tensor, see Appendix \ref{conventions}. \\

One remark is appropriate at this point: if we employed the $d$-dimensional value of the improvement parameter $\chi = (d-2)/[4(d-1)] = 6 - \epsilon/18$,
there should be some extra finite terms in the correlators we compute than when employing directly $\chi = 1/6$, because of the interplay between $\epsilon/18$ 
and the UV pole $\propto 1/\epsilon$, at least in principle. If these terms survived in the $d \rightarrow 4$ limit, then this result would differ from the computation performed directly with $\chi = 1/6$.
This actually happens for individual diagrams. Now, if the limit of a $d$-dimensional correlators were not the $4$d correlators, 
it would mean that conformal correlators depend on the regularization technique employed to compute them in perturbation theory, 
for this is manifestly an issue specific to dimensional regularization (DR hereafter). This is patently inconsistent and suggests that a further consitency condition - beside all the Ward identities we explicitly 
check - is that these extra terms must cancel out when all the pieces making up conformal correlators are put together. 
This turns out to be the case and the explicit check is left to the interested reader.

\subsection{The structure of the correlators: topologies and diagrams} \label{topologies}

Our first step is to exploit the background field method to build the correlation function of $n$ energy-momentum tensors 
and evaluate it diagrammatically at 1 loop. Since the theory is non interacting, as the scalar field is not coupled to any other quantum field, 
there are no higher order perturbative contributions, as one can easily verify.
We work in dimensional regularization DR .

Our definition of the $n-$EMT correlators (also $nT$ below) is that of a symmetric $nth$ functional derivative of $\mathcal{W}$ w.r.t. the metric tensor,
\bea \label{NPF}
< T^{\mu_1\nu_1}(x_1)...T^{\mu_n\nu_n}(x_n)> 
&=&
\bigg[\frac{2^n}{\sqrt{g_{x_1}}\dots \sqrt{g_{x_n}}} \,
\frac{\delta^n \mathcal{W}}{\delta g_{\mu_n\nu_n}(x_n) \ldots \delta g_{\mu_1\nu_1}(x_1)}\bigg]
\bigg|_{g_{\mu\nu} = \delta_{\mu\nu}} \nn \\
&=&
2^n\, \frac{\delta^n \mathcal{W}}{\delta g_{\mu_n\nu_n}(x_n) \ldots \delta g_{\mu_1\nu_1}(x_1)}\bigg|_{g_{\mu\nu} = 
\delta_{\mu\nu}} \, .
\eea
Symmetry comes from leaving factors ${2}/{\sqrt{g}}$ outside of the derivatives.
Given that field theory in Minkowski space is simply given by an analytic continuation from 4d Euclidean theory we work with
through $\delta_{\mu\nu}\rightarrow \eta_{\mu\nu}$, we will also refer to this vertex as to the "n-graviton " vertex.
We choose to denote such correlators, which contain contact terms, with small angular brackets ($< \,\, >$). 

Contact terms are characterized in coordinate space by the presence of at least two gravitons at the same spacetime point. 
Such contact terms are absent by definition in the expression of correlation functions given by the expectation value of the product of $n$ EMT's, 
which are denoted with large angular brackets ($\langle\,\, \rangle$)  as in 
\beq
\langle T^{\mu_1\nu_1}(x_1) \, \ldots \, T^{\mu_n\nu_n}(x_n) \rangle =  \frac{1}{\mathcal N} \int \mathcal D \Phi \, 
T^{\mu_1\nu_1}(x_1) \, \ldots \, T^{\mu_n\nu_n}(x_n)  \, e^{-S} \bigg|_{g_{\mu\nu}=\delta_{\mu\nu}} \,.
\eeq
This distinction will not only apply to the vev of $n$ insertions of the EMT, but also to contact terms and, 
in general, to all the correlation functions appearing in this paper.  

It will also be useful to introduce the following notation to represent the functional derivative 
with respect to the background metric evaluated in flat space,
\bea
\label{funcder}
\left[f(x)\right]^{\muu\nuu\mud\nud\dots\mu_{n}\nu_{n}}(\xu,\xd,\dots,x_n) 
\equiv
\frac{\delta^n\, f(x)}{\delta g_{\mu_n\nu_n}(x_{n}) \, \ldots \, \delta g_{\mud\nud}(\xd) \, \delta g_{\muu\nuu}(\xu)}
\bigg|_{g_{\mu\nu}=\delta_{\mu\nu}} \, .
\eea
Since with our conventions all the functional derivatives will be taken w.r.t. the metric tensor with lower indices, 
which thus produces upper indices, possible functional derivatives with lower indices will mean just
\bea
\left[f(x)\right]_{\muu\nuu\ldots\mu_{n}\nu_{n}}(\xu,\xd,\dots,x_n) 
\equiv
\delta_{\mu_1 \alpha_1} \delta_{\nu_1 \beta_1}\, \ldots\, \delta_{\mu_n \alpha_n}\delta_{\nu_n \beta_n}\,
\left[f(x)\right]^{\alpha_1 \beta_1 \alpha_2 \beta_2 \dots \alpha_{n}\beta_{n}}(\xu,\xd,\dots,x_n)  \,.
\eea
In order to make tensorial expressions more compact, if we happen to contract two indices with the metric, 
we will stack the two contracted indices on top of each other, as in  
\beq 
\left[\mathcal S \right]^{\mu_1}_{\mu_1}\equiv  \left[\mathcal S \right]^{\mu_1\nu_1}\, 
\delta_{\mu_1 \nu_1}\qquad \textrm{or}
\qquad  
\left[\mathcal S \right]^{\mu_1\mu_2}_{\mu_1\mu_2} \equiv  \left[\mathcal S \right]^{\mu_1\nu_1\mu_2\nu_2}
\delta_{\mu_1 \nu_1} \delta_{\mu_2 \nu_2} \, .
\eeq

With these definitions, a single functional derivative of the action in a correlation function is always equivalent, modulo a factor, to an insertion of 
a $T^{\mu\nu}$ in the flat limit, since 
\beq
 \left[\mathcal S \right]^{\mu_1 \nu_1}(x_1) \equiv \frac{\delta \mathcal S}{\delta g_{\mu_1 \nu_1}(x_1)} \bigg|_{g_{\mu\nu}=\delta_{\mu\nu}} =  -\frac{1}{2}T^{\mu_1 \nu_1}(x_1) \, .
 \eeq

\begin{figure}[h]
\begin{center}
\begin{minipage}{0.20\textwidth}
 \centerline{\includegraphics[width=1.5\textwidth]{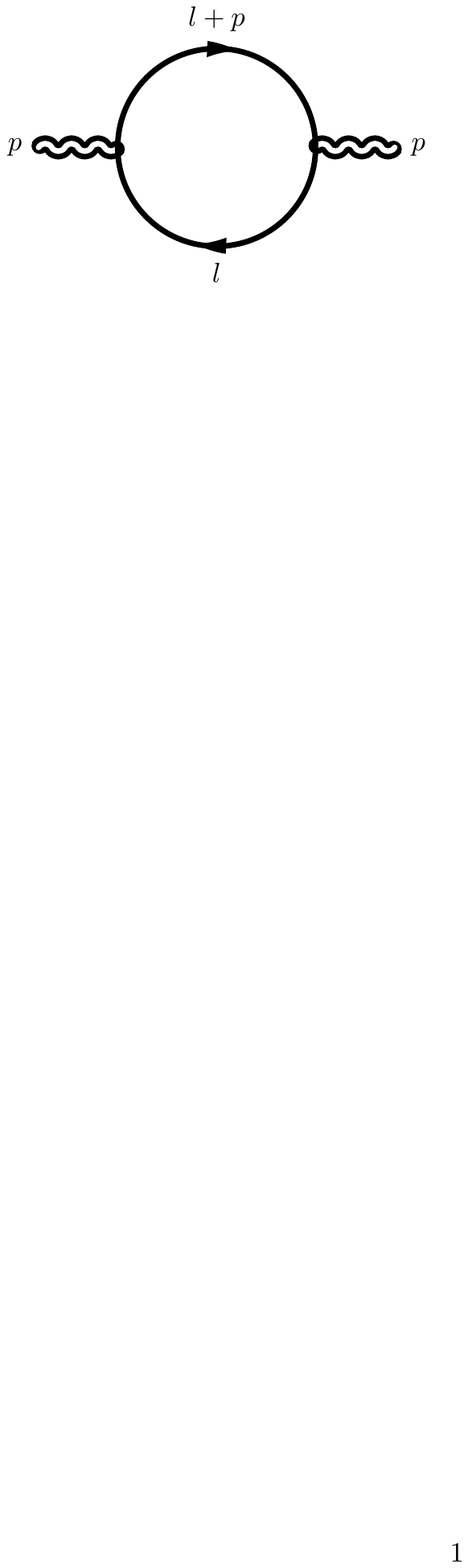}}
\end{minipage}
\caption{The only 1 loop contribution to the two-point correlation function}
\label{t2_topology}
\end{center}
\end{figure}

To begin with the easiest correlation functions, the definition of Eq. (\ref{NPF}) implies that the two-point function is
\bea
< T^{\muu\nuu}(x_1)T^{\mud\nud}(x_2) >
= 4\, \bigg[ \langle \left[\mathcal S \right]^{\muu\nuu}(\xu)\left[\mathcal S \right]^{\mud\nud}(\xd) \rangle - \langle 
[S]^{\muu\nuu\mud\nud}(\xu,\xd) \rangle \bigg] \,.
\eea
The last term on the right hand side of the equation above, which is a massless tadpole, can be set to zero in DR, 
so that the $2T$ correlation function, obtained by differentiation of the generating functional, coincides with the quantum average of two energy-momentum tensors
\bea
< T^{\muu\nuu}(x_1)T^{\mud\nud}(x_2) > &=& 4  \langle \left[\mathcal S \right]^{\muu\nuu}(\xu)\left[\mathcal S 
\right]^{\mud\nud}(\xd) \rangle  
= \langle T^{\muu\nuu}(x_1)T^{\mud\nud}(x_2) \rangle \,.
\eea
This will not be true for higher order correlation functions, where non vanishing contact terms also appear. 
For the sake of completeness, the 1 loop contribution to the two-point correlation function is illustrated in Fig. \ref{t2_topology}.\\

The $3T$ correlator functional expansion is given instead by
\bea
\label{3Tall}
< T^{\muu\nuu}(x_1)T^{\mud\nud}(x_2)T^{\mut\nut}(x_3) >
&=&
\langle T^{\muu\nuu}(\xu) T^{\mud\nud}(\xd)T^{\mut\nut}(\xt)\rangle 
\nn \\
&& \hspace{-75mm} 
-\, 4\, \bigg(\langle \left[\mathcal S\right]^{\muu\nuu\mud\nud}(\xu,\xd)\,T^{\mut\nut}(\xt)\rangle + 2\, \text{perm.} \bigg)
-\, 8\, \langle  \left[\mathcal S\right]^{\muu\nuu\mud\nud\mut\nut}(\xu,\xd,\xt)\rangle
\eea
whose right hand side is expressed in terms of one ordinary three-point correlator plus extra contact terms. 
The additional terms obtained by permutation are such as to render symmetric the right hand side. 

\begin{figure}[h]
\begin{center}
\begin{minipage}{0.20\textwidth}
 \centerline{\includegraphics[width=1.2\textwidth]{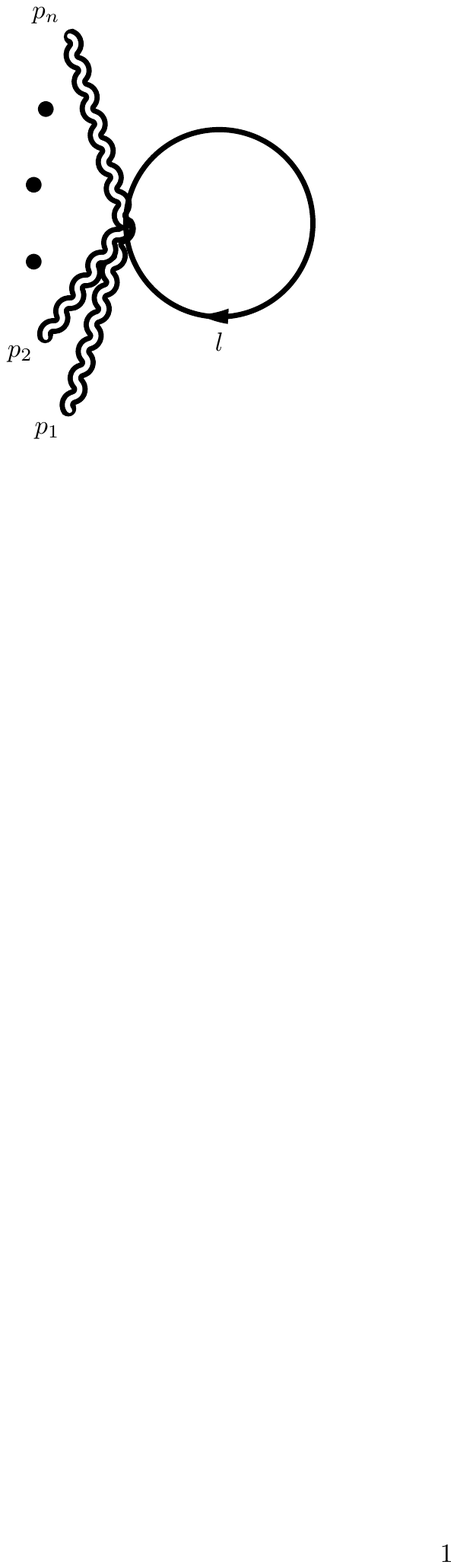}}
\end{minipage}
\caption{The general topology of the vanishing 1-point contributions.}
\label{tadpolesfig}
\end{center}
\end{figure}

\begin{figure}[!h]
\begin{center}
\begin{minipage}{0.20\textwidth}
\centerline{\includegraphics[width=1.5\textwidth]{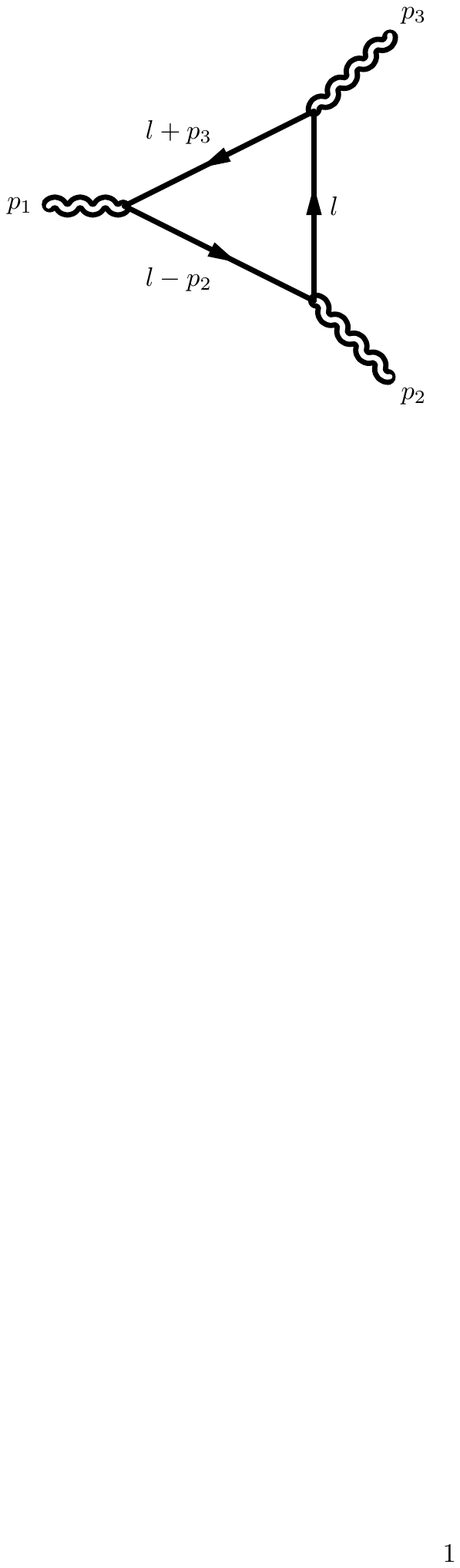}}
\end{minipage}
\hspace{2.5cm}
\begin{minipage}{0.20\textwidth}
\centerline{\includegraphics[width=1.5\textwidth]{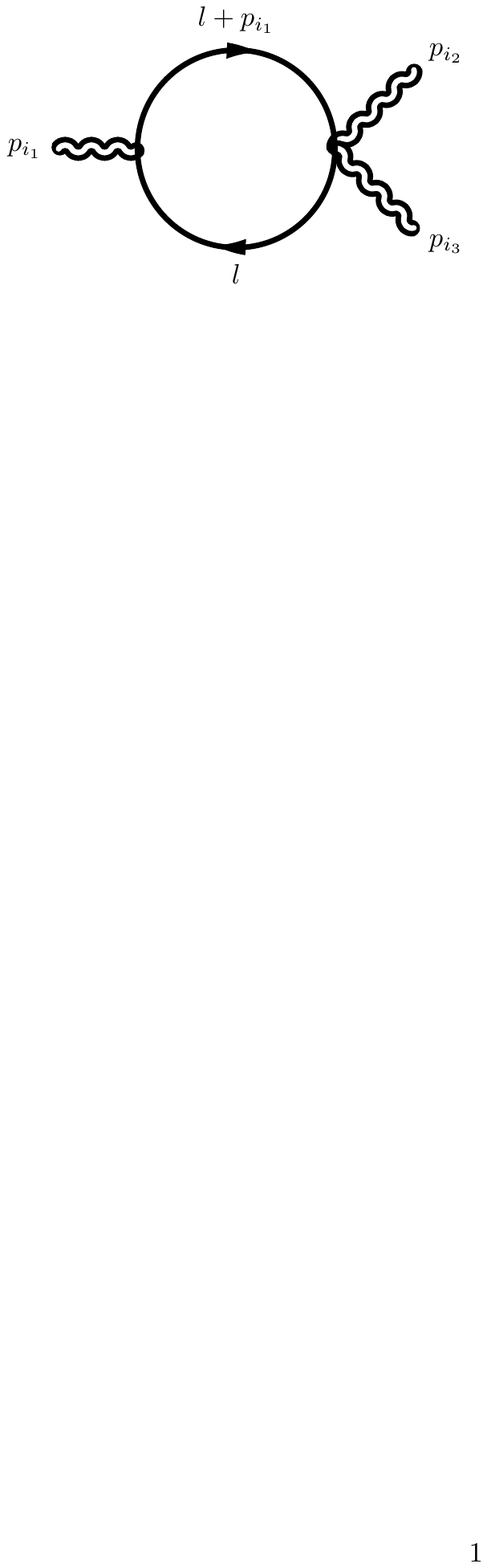}}
\end{minipage}
\caption{The topologies contributing to the  $<TTT>$ correlation function.
All the external momenta are incoming.}
\label{topologies_ttt}
\end{center}
\end{figure}

The first term on the right hand side of Eq. (\ref{3Tall}) is an ordinary three-point function, 
whose connected component is given, at 1 loop, by the triangle diagram of Fig. \ref{topologies_ttt}, 
while the last term is a massless tadpole (see Fig. \ref{tadpolesfig})\footnote{All Feyman diagrams were drawn with~\emph{feynMF}.~\cite{Ohl:1995kr}}, 
which can be set to zero
\beq
\langle \left[\mathcal S\right]^{\muu\nuu\mud\nud\mut\nut}(\xu,\xd,\xt)\rangle=0. 
\eeq
In the $3T$ case, contact terms have the topology of a bubble and are generated by correlators containing insertions of the second 
functional derivative of the action w.r.t. to the metric. 
Their general structure is shown in Fig. \ref{topologies_ttt} and each one of them is simply obtained by 
assigning the momenta to the diagram according to one of the the three possible groupings. \\

\begin{figure}[!h]
\begin{center}
\begin{minipage}{0.20\textwidth}
 \centerline{\includegraphics[width=1.5\textwidth]{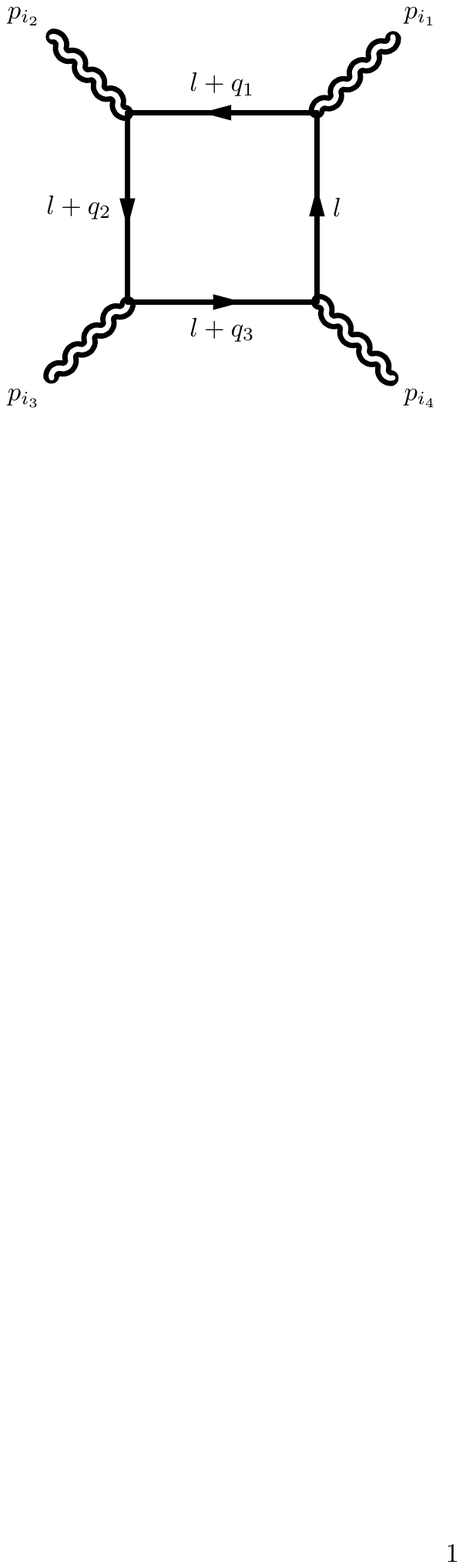}}
\end{minipage}
\hspace{2.5cm}
\begin{minipage}{0.20\textwidth}
\centerline{\includegraphics[width=1.5\textwidth]{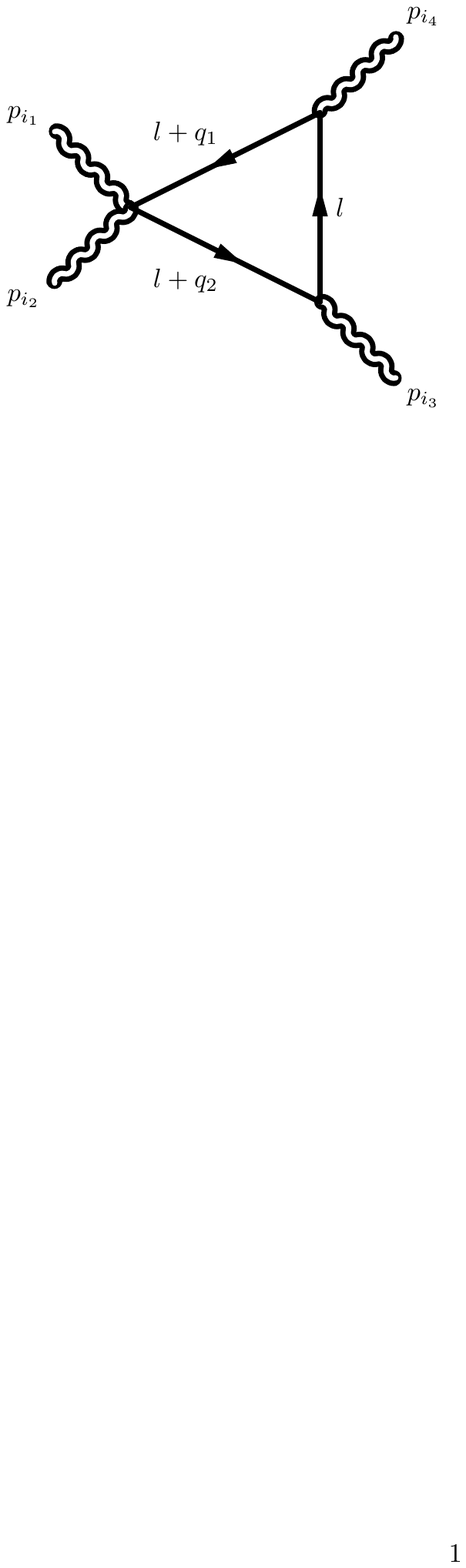}}
\end{minipage}
\caption{The square and triangle topologies contributing to the  $<TTTT>$ correlation function.
All the external momenta are incoming.}
\label{4T_topologies_1}
\end{center}
\end{figure}

\begin{figure}[!h]
\begin{center}%
\begin{minipage}{0.20\textwidth}
\centerline{\includegraphics[width=1.5\textwidth]{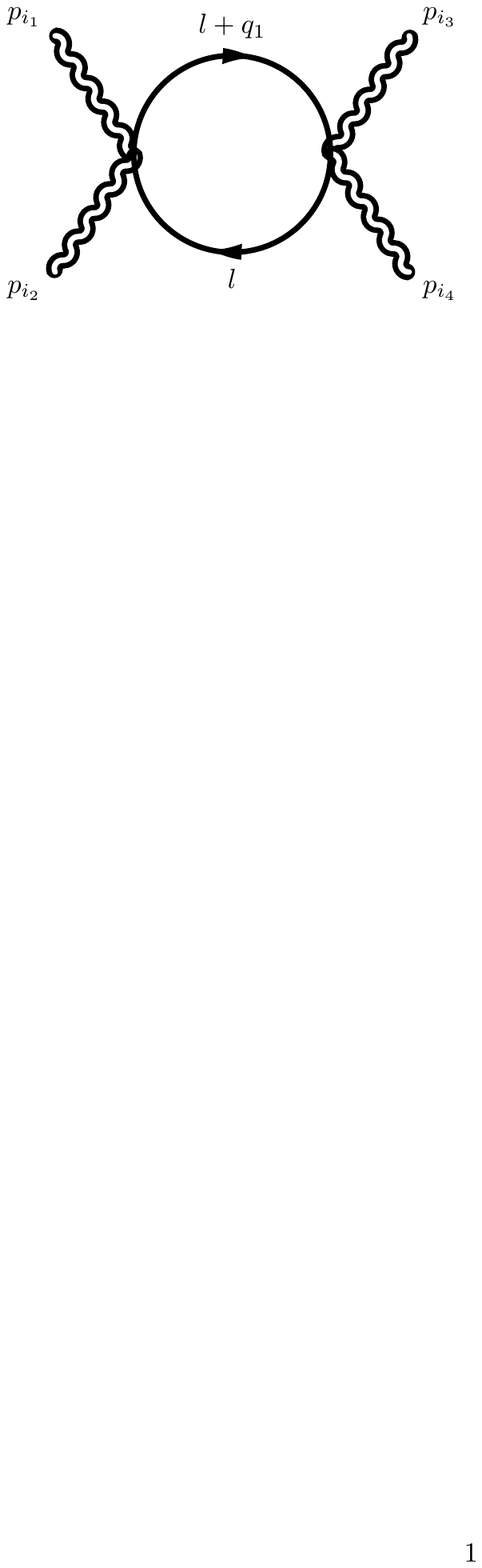}}
\end{minipage}
\hspace{2.5cm}%
\begin{minipage}{0.20\textwidth}
\centerline{\includegraphics[width=1.5\textwidth]{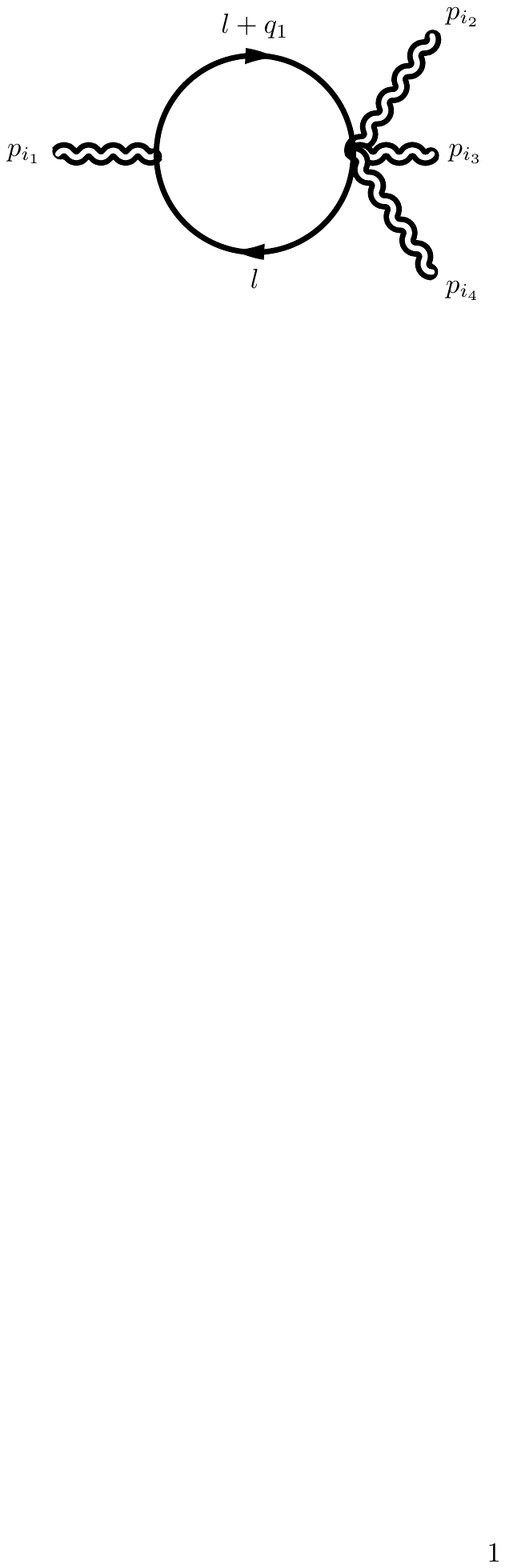}}
\end{minipage}
\caption{
The two kinds of bubble topology contributing to the  $<TTTT>$ correlation function in a scaleless theory.
All the external momenta are incoming.}
\label{4T_topologies_2}
\end{center}
\end{figure}

Moving finally to the $4T$ case, a similar expansion holds and is given by
\bea
< T^{\muu\nuu}(x_1)T^{\mud\nud}(x_2)T^{\mut\nut}(x_3)T^{\muq\nuq}(x_4)>
&=&
\langle T^{\muu\nuu}(x_1)T^{\mud\nud}(x_2)T^{\mut\nut}(x_3)T^{\muq\nuq}(x_4)\rangle
\nn \\
&& \hspace{-75 mm}
-\, 4\, \bigg[\langle \left[\mathcal S\right]^{\muu\nuu\mud\nud}(\xu,\xd)T^{\mut\nut}(\xt) T^{\muq\nuq}(\xq)\rangle  + 5 \, \text{perm.} \bigg] 
\nn \\
&& \hspace{-75mm}
+\, 16\, \bigg[\langle \left[\mathcal S\right]^{\muu\nuu\mud\nud}(\xu,\xd)\left[\mathcal S\right]^{\mut\nut\muq\nuq}(\xt,\xq)\rangle + 2 \, \text{perm.} \bigg]
\nn \\
&& \hspace{-75mm}
- \, 8\, \bigg[\langle \left[\mathcal S\right]^{\muu\nuu\mud\nud\mut\nut}(\xu,\xd,\xt)T^{\muq\nuq}(\xq)\rangle + 3 \, \text{perm.} \bigg]
- \, 16\, \langle \left[\mathcal S\right]^{\muu\nuu\mud\nud\mut\nut\muq\nuq}(\xu,\xd,\xt,\xq)\rangle
\nn \\
\label{4PF}
\eea
with 
\beq
\langle \left[\mathcal S\right]^{\muu\nuu\mud\nud\mut\nut\muq\nuq}(\xu,\xd,\xt,\xq)\rangle=0, 
\eeq
being a massless tadpole contribution. In this case the perturbative expansion generates three diagrams of box type, represented by the Green function with 4 EMT insertions 
on the right hand side of (\ref{4PF}), plus triangle, bubble and tadpole diagrams generated by the contact terms and graphically represented in Figs. \ref{4T_topologies_1} and \ref{4T_topologies_2}.
The analysis of these contributions is more involved compared to the $3T$ case.
In Figs. \ref{4T_topologies_1} and \ref{4T_topologies_2} we illustrate the general structures of the four kinds of diagrams involved. 
The momenta running through them in addition to the loop momentum $l$ must be replaced by the specific assignments detailed below. 
Later, when we illustrate the implementation of the computation in \wolfram, we will refer to these basic diagrams corresponding to 
inequivalent topologies as \emph{blueprint diagrams}.

For each topology we have a different number of contributions,
corresponding to cyclically inequivalent orderings of the external momenta.

For the square topology, the 3 distinct contributions can be parametrized, for example,
by the following three assignments of momenta (compare to the first diagram in Fig. \ref{4T_topologies_1})
\bea
&&
\left(  p_{i_1}, p_{i_2}, p_{i_3}, p_{i_4} \right) =
\left\{ \begin{array}{l}
\left( p_2, p_3, p_4, p_1 \right)
\\
\left( p_3, p_4, p_2, p_1 \right)
\\
\left( p_4, p_2, p_3, p_1 \right)
\end{array} \right.
\nn \\
&&
q_1 = p_{i_1} \, , \quad q_2 = p_{i_1} + p_{i_2} \, , \quad q_3 = p_{i_1} + p_{i_2} + p_{i_3} = - p_{i_4} 
\label{distinct_square}
\eea

For the triangle topology, there are the following 6 distinct contributions
\bea
&&
\left(  p_{i_1}, p_{i_2}, p_{i_3}, p_{i_4} \right) =
\left\{ \begin{array}{l}
\left( p_1, p_2, p_3, p_4 \right)
\\
\left( p_1, p_3, p_2, p_4 \right)
\\
\left( p_1, p_4, p_2, p_3 \right)
\\
\left( p_2, p_3, p_1, p_4 \right)
\\
\left( p_2, p_4, p_1, p_3 \right)
\\
\left( p_3, p_4, p_1, p_2 \right)
\end{array} \right.
\nn \\
&&
q_1 = p_{i_4} \, , \quad q_2 = - p_{i_3}  \, .
\label{distinct_triangle}
\eea

As for the two bubble topologies, we choose to distinguish them through the nomenclature of $22$-bubble and $31$-bubble, 
the figures standing for the numbers of gravitons meeting in each of their respective vertices. 

The 3 contributions for the $22$-bubble are
\bea
&&
\left(  p_{i_1}, p_{i_2}, p_{i_3}, p_{i_4} \right) =
\left\{ \begin{array}{l}
\left( p_1, p_2, p_3, p_4 \right)
\\
\left( p_1, p_3, p_2, p_4 \right)
\\
\left( p_1, p_4, p_2, p_3 \right)
\end{array} \right.
\nn \\
&&
q_1 = p_{i_1} + p_{i_2} \, .
\label{distinct_22bubble}
\eea
The $4$ inequivalent contributions for the $31$-bubbles are instead
\bea
&&
\left(  p_{i_1}, p_{i_2}, p_{i_3}, p_{i_4} \right) =
\left\{ \begin{array}{l}
\left( p_1, p_2, p_3, p_4 \right)
\\
\left( p_2, p_1, p_3, p_4 \right)
\\
\left( p_3, p_1, p_2, p_4 \right)
\\
\left( p_4, p_1, p_2, p_3 \right)
\end{array} \right.
\nn \\
&&
q_1 = p_{i_1} \, . 
\label{distinct_22bubble}
\eea
The task of computing the three-point function in a completely off-shell configuration and checking at the same time the 
transverse and trace Ward identities was already performed in~\cite{Coriano:2012wp}, but only the result in a partially on-shell
configuration was explicitly given. Now we face a more demanding task, which is the computation of the four-point function in a completely
off-shell kinematics. This was made possible by the development of \patel~\cite{Patel:2016fam}, a \wolfram package for tensor algebra and automatic reduction
of 1 loop tensor integrals of any rank in arbitrary even dimensions. 
We detail the computation in section \ref{mathematica}. For now, we just mention that also the step-by-step 
computations of the two and (much more significanly) the three-point functions are made available, 
so as to give the reader a full overview of the automated procedure in simpler setups, beside the target case of the $4T$.  

The building blocks of our computation are the scalar-graviton interaction vertices, which we report in Appendix \ref{Vertices}.
Although we were completely confident about their calculation and we tested \patel extensively by reproducing 
all of our previous results of~\cite{Coriano:2012wp} and a few other field theory results, the reduction of rank-8 tensor integrals still was an uncharted territory.
Just as it was the case for the $3T$, we checked all of our computations by testing the transverse and trace 
Ward identities our four-point correlator was supposed to satisfy, which we derive in section \ref{ward}.

\subsection{Organization of the \wolfram files}

This is a quick overview of the files developed in order to perform and test our calculation of the 4-point function.
The same picture is given by the README file stored in our repository.
Some of the calculations can be quite time consuming and the numerical checks of the Ward identities for 
the 4 point function requires your computer to have at least 25 GB of memory at its disposal in order not to crush. \\

{\bf Some necessary requirements on the machine}: \wolfram (version 10.1 or higher); the additional packages \patel and \denner 
must be loaded on the machine where the notebooks are run. Beside, the notebook \emph{functional\_derivatives.nb} must be run once in order to be able to run the rest of the calculation, 
particularly \emph{correlators\_calculation.nb}, which must in turn be run in order to generate the correlation functions which are 
checked by \emph{ward\_identities.nb}.\\

Here follows a concise description of the scope and purpose of each notebook: 
\begin{itemize}
\item The notebook \emph{tensor\_bases/tensor\_bases\_generation.nb} generates the 4 files \emph{tensmom\#rank\#\#}. 
As the name suggests, the tensors in each of these files  span a complete basis of tensors which are rank-\#\# products of the metric tensor and \# independent momenta. 
They are needed to check the Ward identities for the three and four-point correlation functions.
\item \emph{functional\_derivatives.nb} generates the file \emph{all\_functional\_derivatives}. 
   As detailed in the paper, our computation requires heavy use of tensor strutures: rank-2, 4 and 6 trace anomalies
   for the two, three and four-point functions, rank-4, 6 and 8 countertems for the very same correlators. 
   All of them are obtained by functionally differentiating scalars consisting of algebraic combinations of the Riemann tensor, the Ricci tensor and the Ricci scalar.
   The notebook starts by explaining the simplest functional derivatives of the metric tensor and goes on all the way up
   to anomalies, counterterms and interaction vertices of the scalar with the background gravitational field, introducing gradually more complex structures. 
   The second part of the notebook checks that the counterterms and the anomalies, which are computed non
   perturbatively, obey all the constraints they are supposed to. This is needed to make us more confident about our 
   explicit calculations of the Green functions, provided that their divergent parts match the counterterms (they do indeed) 
   and that they pass the check of the trace Ward identities with the anomalies (they do as well).    
\item \emph{correlators\_calculation.nb} explicitly computes the two, three and four-point functions, checks that they match the counterterms 
  computed in \emph{functional\_derivatives.nb} and stores them in 3 files: \emph{T2\_scalar}, \emph{T3\_scalar} and \emph{T4\_scalar}.
\item The .jpg files in the "figures" folder are simply graphical representations of the diagrams computed in    
  \emph{correlators\_calculation.nb} and of the vertices computed in \emph{functional\_derivatives.nb}, 
  which are loaded in the same notebooks just above the line of code computing each of them.
\item The notebook \emph{ward\_identities.nb} checks the Ward identities for all of our correlation functions; analytically for two and three-point, numerically for four-point.
\item The notebook \emph{vanishing\_euler\_ct.nb} proves that the Euler counterterm for the 4-point function actually vanishes in 4d (see section \ref{counterterm}).
\end{itemize}

\section{Derivation of the transverse and trace Ward identities}\label{ward}%

In this section, we derive the Ward identities stemming form the requirements of invariance of the generating functional
under general diffeomorphisms and Weyl transformations. We call them transverse and trace Ward identities respectively.   
We proceed with a derivation of the relevant Ward identities satisfied by the two, three and four-point functions of the EMT.\\

Invariance under diffeomorphisms is defined by the condition of general covariance of the generating functional $\mathcal W[g]$, which translates into 
\bea
\nabla_{\nuu} < T^{\muu\nuu}(x_1) >_g
&=& 
\nabla_{\nuu} \bigg(\frac{2}{\sqrt{g_{x_1}}}\frac{\delta\mathcal W[g]}{\delta g_{\muu\nuu}(x_1)}\bigg)
\nn \\
&=&
\pd_{\nuu} < T^{\muu\nuu}(x_1) >_g - \Gamma^{\muu}_{\kappa\nuu}\, < T^{\kappa\nuu}(x_1) >_g - \Gamma^{\nuu}_{\nuu\kappa}\, < T^{\muu\kappa}(x_1) >_g
\nn \\
&\Rightarrow&
2\,\pd_{\nuu} < \frac{\delta \mathcal W}{\delta g_{\muu\nuu}(x)} >_g - \Gamma^{\muu}_{\kappa\nuu}\, < \frac{\delta\mathcal W}{\delta g_{\kappa\nuu}(\xu)} >_g = 0  \, , 
\label{masterward}
\eea
where in the last step, crucially for the symmetry of the correlators in the derived Ward identities below, we have exploited
the cancellation of the last term on the rhs of the second line with the derivative of $1/\sqrt{g_{\xu}}$.

The Ward identities for symmetric correlators we are after are obtained by functional differentiation 
of Eq. (\ref{masterward}) as many times as it takes to get the correlator we are interested in, followed by taking the flat limit. \\

For the $2T$ we have the well known transverse condition,
\bea
\pd_{\nuu} < T^{\muu\nuu}(x_1) T^{\mud\nud}(x_2) > &=& 0 \, . \label{WI2PFCoordinateFlat}
\eea

For the $3T$ we see an already non trivial rhs showing up
\bea
\pd_{\nuu} < T^{\muu\nuu}(x_1)T^{\mud\nud}(x_2)T^{\mut\nut}(x_3) > 
&=& 
- 2\, \left[\Gamma^{\muu}_{\kappa\nuu}(x_1)\right]^{\mud\nud}(x_2) \langle T^{\kappa\nuu}(x_1)T^{\mut\nut}(x_3) \rangle
\nn \\
&&
- 2\, \left[\Gamma^{\muu}_{\kappa\nuu}(x_1)\right]^{\mut\nut}(x_3) \langle T^{\kappa\nuu}(x_1)T^{\mud\nud}(x_2) \rangle \, .
\label{WI3PFCoordinateFlat} 
\eea

Much more involved is the case for the $4T$, whose transverse Ward identity in coordinate space is given by
\bea
\pd_{\nuu} < T^{\muu\nuu}(x_1)T^{\mud\nud}(x_2)T^{\mut\nut}(x_3)T^{\muq\nuq}(x_4) > 
&=&
\nn \\
&& \hspace{-55mm}
- 2\, \bigg[\left[\Gamma^{\muu}_{\kappa\nuu}(x_1)\right]^{\mud\nud}(x_2)
 < T^{\kappa\nuu}(x_1) T^{\mut\nut}(x_3)T^{\muq\nuq}(x_4) > 
\nn \\
&& \hspace{-55mm}
+ 2\, \left[\Gamma^{\muu}_{\kappa\nuu}(x_1)\right]^{\mut\nut\muq\nuq}(\xt,\xq)
 \langle T^{\kappa\nuu}(x_1)T^{\mud\nud}(\xd) \rangle 
+ \big( 2 \leftrightarrow 3, 2 \leftrightarrow 4 \big) \bigg] \,  .
\label{WI4PFCoordinateFlat} 
\eea
The functional derivatives of the Christoffel symbol, obtained from the expansion of the covariant derivatives which appear in previous equations, are explicitly given in Appendix \ref{Vertices}. \\

Before moving to momentum space, we point out that the Fourier transform formula is defined by (\ref{fourier}), where all the momenta are incoming. 
We keep this convention throughout the paper and in all the momentum space computations in our code. \\
By applying(\ref{fourier}) and some integrations by parts to our Ward identities, we get the momentum space transverse Ward identities, 
which are a set of algebraic contraints. 

The momentum space $2T$ satisfies 

\beq \label{transverse2t}
p_{\nuu} \langle T^{\muu\nuu}(p)T^{\mud\nud}(-p) \rangle = 0.
\eeq

For the three and four-point functions, the transverse Ward identities are quite cumbersome to write down fully expanded. 

We present an expanded version of the first transverse Ward identity for the thee-point function for illustration, whereas we keep
the full sets given below implicit and refer the reader to Appendix \ref{conventions} for a list of the explicit momentum space forms of the constituting elements
\bea
p_{1\,\nuu} < T^{\muu\nuu}(\pu) T^{\mud\nud}(\pdd) T^{\mut\nut}(\pt) >
&=&
- p_3^{\muu} \langle T^{\mut\nut}(\pdd)  T^{\mud\nud}(- \pdd)  \rangle 
- p_2^{\muu} \langle T^{\mud\nud}(\pt) T^{\mut\nut}(- \pt) \rangle \nn \\
&&  \hspace{-3cm}
+ p_{3\, \nuu} \bigg[\delta^{\muu\nut} \langle T^{\nuu\mut}(\pdd) T^{\mud\nud}(-\pdd) \rangle
+ \delta^{\muu\mut} \langle T^{\nuu\nut}(\pdd) T^{\mud\nud}(-\pdd)  \rangle \bigg] 
\nn \\
&&  \hspace{-3cm}
+ p_{2\, \nuu} \bigg[\delta^{\muu\nud} \langle T^{\nuu\mud}(\pt) T^{\mut\nut}(-\pt)  \rangle
+ \delta^{\muu\mud}  \langle T^{\nuu\nud}(\pt) T^{\mut\nut}(-\pt) \rangle\bigg] \, .
\eea

For every $nT$, there is of course one such Ward identity for each EMT.
We present the full set here below for the three-point function in compact form,
\bea \label{transverse3t}
p_{1\,\nuu} < T^{\muu\nuu}(\pu) T^{\mud\nud}(\pdd) T^{\mut\nut}(\pt) >
&=&
- 2\,i\, \left[\Gamma^{\muu}_{\kappa\nuu}\right]^{\mud\nud}(\pdd) \langle T^{\kappa\nuu}(\pt)T^{\mut\nut}(-\pt) \rangle + (2\leftrightarrow 3)
\nn \\
p_{2\,\nud} < T^{\muu\nuu}(\pu) T^{\mud\nud}(\pdd) T^{\mut\nut}(\pt) >
&=&
- 2\,i\, \left[\Gamma^{\mud}_{\kappa\nud}\right]^{\muu\nuu}(\pu) \langle T^{\kappa\nud}(\pt)T^{\mut\nut}(-\pt) \rangle + (1\leftrightarrow 3)
\nn \\
p_{3\,\nut} < T^{\muu\nuu}(\pu) T^{\mud\nud}(\pdd) T^{\mut\nut}(\pt) >
&=&
- 2\,i\, \left[\Gamma^{\mut}_{\kappa\nut}\right]^{\muu\nuu}(\pu) \langle T^{\kappa\nut}(\pdd)T^{\mud\nud}(-\pdd) \rangle + (1\leftrightarrow 2) \, .
\nn \\
\eea

Finally, for the four-point correlator we have 4 transverse Ward identities (this is the last set for which we report all channels; in what
follows only one channel for every Ward identity will be explicitly written, though all of them were tested),
\bea \label{transverse4t}
&&\hspace{-20mm}
p_{1\,\nuu} \,< T^{\muu\nuu}(\pu)T^{\mud\nud}(\pdd)T^{\mut\nut}(\pt)T^{\muq\nuq}(\pq) > =
\nn \\
&&
2\,i\, \bigg[\left[\Gamma^{\muu}_{\kappa\nuu}\right]^{\mud\nud}(\pdd) < T^{\kappa\nuu}(-\pt-\pq) T^{\mut\nut}(\pt)T^{\muq\nuq}(\pq) > 
\nn \\
&&
-2\, \left[\Gamma^{\muu}_{\kappa\nuu}\right]^{\mut\nut\muq\nuq}(\pt,\pq)  \langle T^{\kappa\nuu}(-\pdd)T^{\mud\nud}(\pdd) \rangle 
+ \big( 2 \leftrightarrow 3, 2 \leftrightarrow 4 \big) \bigg] \, , 
\nn \\
&&\hspace{-20mm}
p_{2\,\nu2} \,< T^{\muu\nuu}(\pu)T^{\mud\nud}(\pdd)T^{\mut\nut}(\pt)T^{\muq\nuq}(\pq) > =
\nn \\
&&
2\,i\, \bigg[\left[\Gamma^{\mud}_{\kappa\nud}\right]^{\muu\nuu}(\pu) < T^{\kappa\nud}(-\pt-\pq) T^{\mut\nut}(\pt)T^{\muq\nuq}(\pq) > 
\nn \\
&&
-2\, \left[\Gamma^{\mud}_{\kappa\nud}\right]^{\mut\nut\muq\nuq}(\pt,\pq)  \langle T^{\kappa\nud}(-\pu)T^{\muu\nuu}(\pu) \rangle 
+ \big( 1 \leftrightarrow 3, 1 \leftrightarrow 4 \big) \bigg] \, , 
\nn \\
&& \hspace{-20mm}
p_{3\,\nu3} \,< T^{\muu\nuu}(\pu)T^{\mud\nud}(\pdd)T^{\mut\nut}(\pt)T^{\muq\nuq}(\pq) > =
\nn \\
&&
2\,i\, \bigg[\left[\Gamma^{\mut}_{\kappa\nut}\right]^{\muu\nuu}(\pu) < T^{\kappa\nut}(-\pdd-\pq) T^{\mud\nud}(\pdd)T^{\muq\nuq}(\pq) > 
\nn \\
&&
-2\, \left[\Gamma^{\mut}_{\kappa\nut}\right]^{\mud\nud\muq\nuq}(\pdd,\pq)  \langle T^{\kappa\nut}(-\pu)T^{\muu\nuu}(\pu) \rangle 
+ \big( 1 \leftrightarrow 2, 1 \leftrightarrow 4 \big) \bigg] \, , 
\nn \\
&&\hspace{-20mm}
p_{4\,\nu4} \,< T^{\muu\nuu}(\pu)T^{\mud\nud}(\pdd)T^{\mut\nut}(\pt)T^{\muq\nuq}(\pq) > = 
\nn \\
&&
2\,i\, \bigg[\left[\Gamma^{\muq}_{\kappa\nuq}\right]^{\muu\nuu}(\pu) < T^{\kappa\nuq}(-\pdd-\pt) T^{\mud\nud}(\pdd)T^{\mut\nut}(\pt) > 
\nn \\
&&
-2\, \left[\Gamma^{\muq}_{\kappa\nuq}\right]^{\mud\nud\mut\nut}(\pdd,\pt)  \langle T^{\kappa\nuq}(-\pu)T^{\muu\nuu}(\pu) \rangle 
+ \big( 1 \leftrightarrow 2, 1 \leftrightarrow 3 \big) \bigg] \, .
\label{WI4PFMomentumFlat}
\eea

Deriving the trace Ward identities is simpler. Rewriting (\ref{traceanomaly}) with $n_I =1$ as
\beq
g_{\muu\nuu}\, \frac{\delta \mathcal W}{\delta g_{\muu\nuu}(\xu)} = \frac{\sqrt{g}}{2}\, \mathcal A[g] \, , 
\eeq
we can functionally differentiate up to three more times and obtain, 
after Fourier-transforming to momentum space with (\ref{fourier}),
\beq
\delta_{\muu\nuu} < T^{\muu\nuu}(-\pu) T^{\mud\nud}(\pu) > = \left[ \sqrt{g}, \mathcal A[g]\right]^{\mud\nud}(\pu) \, , 
\label{trace2t}
\eeq
\bea
\delta_{\muu\nuu} < T^{\muu\nuu}(\pu) T^{\mud\nud}(\pdd)T^{\mut\nut}(\pt) > 
&=& 
\left[ \sqrt{g} \, \mathcal A[g]\right]^{\mud\nud\mut\nut}(\pdd,\pt) \, , 
\nn \\
&& \hspace{-20mm}  
- < T^{\mud\nud}(-\pdd)T^{\mut\nut}(\pdd) > - < T^{\mud\nud}(-\pt)T^{\mut\nut}(\pt) > 
\label{trace3t}
\eea
\bea
\delta_{\muu\nuu} < T^{\muu\nuu}(\pu) T^{\mud\nud}(\pdd)T^{\mut\nut}(\pt)T^{\muq\nuq}(\pq) > 
&=& 
\left[ \sqrt{g} \, \mathcal A[g]\right]^{\mud\nud\mut\nut\muq\nuq}(\pdd,\pt,\pq) \, , 
\nn \\
&& \hspace{-20mm}
- < T^{\mud\nud}(-\pt-\pq)T^{\mut\nut}(\pt)T^{\muq\nuq}(\pq) > 
\nn \\
&& \hspace{-20mm} 
- < T^{\mut\nut}(-\pdd-\pq)T^{\mud\nud}(\pdd)T^{\muq\nuq}(\pq) > 
\nn \\
&& \hspace{-20mm} 
- < T^{\muq\nuq}(-\pdd-\pt)T^{\mud\nud}(\pdd)T^{\mut\nut}(\pt) > \, .
\label{trace4t}
\eea
The structure of the anomalies for our correlators is implicitly given by
\bea
\bigg[ \mathcal A[g] \bigg]^{\mud\nud}(\pdd) &=& -\frac{2}{3}\,\beta_a\, \left[ \square R \right]^{\mud\nud}(\pdd) \, , 
\nn \\
\bigg[ \mathcal A[g] \bigg]^{\mud\nud\mut\nut}(\pdd,\pt) &=& \beta_a\, \bigg( \left[F\right]^{\mud\nud\mut\nut}(\pdd,\pt) -\frac{2}{3}\, \left[ \sqrt{g}\square R \right]^{\mud\nud\mut\nut}(\pdd,\pt)\bigg) 
\nn \\
&+& 
\beta_b \, \left[G\right]^{\mud\nud\mut\nut}(\pdd,\pt) \, , 
\nn \\
\bigg[ \mathcal A[g] \bigg]^{\mud\nud\mut\nut\muq\nuq}(\pdd,\pt,\pq) &=& 
\beta_a\, \bigg( \left[ \sqrt{g}\,F\right]^{\mud\nud\mut\nut\muq\nuq}(\pdd,\pt,\pq) -\frac{2}{3}\, \left[ \sqrt{g}\, \square R \right]^{\mud\nud\mut\nut\muq\nuq}(\pdd,\pt,\pq)\bigg) 
\nn \\
&+& \beta_b \, \left[\sqrt{g}\, G\right]^{\mud\nud\mut\nut\muq\nuq}(\pdd,\pt,\pq) \, , 
\label{Anomalies234}
\eea
whereas the explicit construction is available in the notebook \emph{functional\_derivatives.nb}.
In the next section we discuss in detail their connection with 1 loop counterterms and how we can use the two for a preliminary 
test of our calculation of the four-point correlator with \patel.

One comment about the anomaly is in order: since the term $\propto \square R$ can be removed by either 
adding an integral $\propto \int d^dx R^2$~\cite{Capper:1974ic} or by promoting the numerical coefficients in the Weyl counterterm to be functions of 
the spacetime dimension $d$~\cite{Osborn:1993cr}, many an author choose to do so. 
Elsewhere we have discussed these issues in detail too~\cite{Coriano:2012wp}, but we will not linger over it here. 
In this paper, we just perform the calculation in DR with no supplemental local counterterm, 
so that (\ref{traceanomalyfinal}) is the correct generating functional for the trace anomaly of our correlators.  

Finally, there are the Ward identities implied by special conformal transformations, but we will not deal with them in the present work.
For a detailed discussion of the derivation and the solution of these identities for two and three-point functions, see~\cite{Bzowski:2013sza}. 

\section{Counterterms, anomalies and a preliminary test of the $4T$ correlator}\label{counterterm}

In this section we review the derivation of the 1 loop couterterms for EMT correlators and we derive the Ward
identities which they must satisfy and which connect them with the respective trace anomalies. 
We explicitly compute these counterterms and check all of the mentioned Ward identities.
Since both the counterterms and the anomalies for the $4T$ correlator 
are highly non trivial structures, the successful check of these tests is a very strong indicator that their calculation is correct.

Therefore, if it is possible to evaluate only the ultraviolet pole of our four-point function, after putting together the
tensor integrals corresponding to the diagrams of section \ref{setup}, we can compare it to the independently 
tested counterterm. If the two expressions match, we have a strong preliminary hint that the diagrams have been 
assembled correctly. We actually performed this test, as the \emph{LoopIntegrate} routine of \patel, 
which substitutes explicit expressions of scalar coefficients into tensor integrals, has an option for
extracting only the ultraviolet pole of each scalar coefficient. 
Summing up the poles of all diagrams, we did match the rank-8 counterterm for our correlator. \\

In order to be as self-contained as possible, we provide here the general formulas for the Passarino-Veltman coefficient functions we employ~\cite{Passarino:1978jh},
so as to clarify the meaning of the symbols $PVB, PVC$ and $PVD$ that the reader will encounter in the snippets of code to follow and in the notebooks.
In the following formula, the symbol $r$ stands for the number of times the metric tensor enters in the symmetric tensor multiplying the coefficient function, 
whereas $n_1$, $n_2$ and $n_3$ stand for the number of times each of the external momenta enters in it,  
\bea
&&
PVB[r,n_1,s,m_0,m_1]  \equiv
\mathbf{B}_{\underbrace{0...0}_{2r} \underbrace{1\dots 1}_{n_1}} \, , \, \text{coefficient of} \, 
\left\{ \left[p_1\right]^{n_1}\, \left[g\right]^r  \right\}^{\muu\dots\mu_{2r+n_1}} \, ,
\nn \\
&&
PVC[r,n_1,n_2,s_1,s_{12},s_2,m_0,m_1,m_2]  \equiv
\mathbf{C}_{\underbrace{0...0}_{2r} \underbrace{1\dots 1}_{n_1}\underbrace{2 \dots 2}_{n_2}}\, , \,
\nn \\
&&\hspace{55mm}
\text{coefficient of}\, \left\{ \left[p_1\right]^{n_1}\, \left[p_2\right]^{n_2}\, \left[g\right]^r  \right\}^{\muu\dots\mu_{2r+n_1+n_2}} \, ,
\nn \\
&&
PVD[r,n_1,n_2,n_3,s_1,s_2,s_3,s_4,s_{12},s_{23},m_0,m_1,m_2,m_3]  \equiv
\mathbf{D}_{\underbrace{0...0}_{2r} \underbrace{1\dots 1}_{n_1}\underbrace{2 \dots 2}_{n_2}\underbrace{3 \dots 3}_{n_3}} \, , \,
\nn \\
&& \hspace{55mm}
\text{coefficient of} \left\{ \left[p_1\right]^{n_1}\, \left[p_2\right]^{n_2}\,\left[p_3\right]^{n_3}\, \left[g\right]^r  \right\}^{\muu\dots\mu_{2r+n_1+n_2+n_3}}\, .
\label{PVreduce}
\eea

The notebook \emph{correlators\_calculation.nb} presents the procedure for all of the EMT correlators through four-point, 
giving also plenty of details about the calculation of the whole correlators.\footnote{
In the conventions of \patel, one works in the $\overline{MS}$ scheme and must think of $1/\epsilon$ as of $1/\epsilon - \gamma + \log(4\pi)$, $\gamma$ being the Euler-Mascheroni constant.} \\

\begin{figure}[h]
\includegraphics{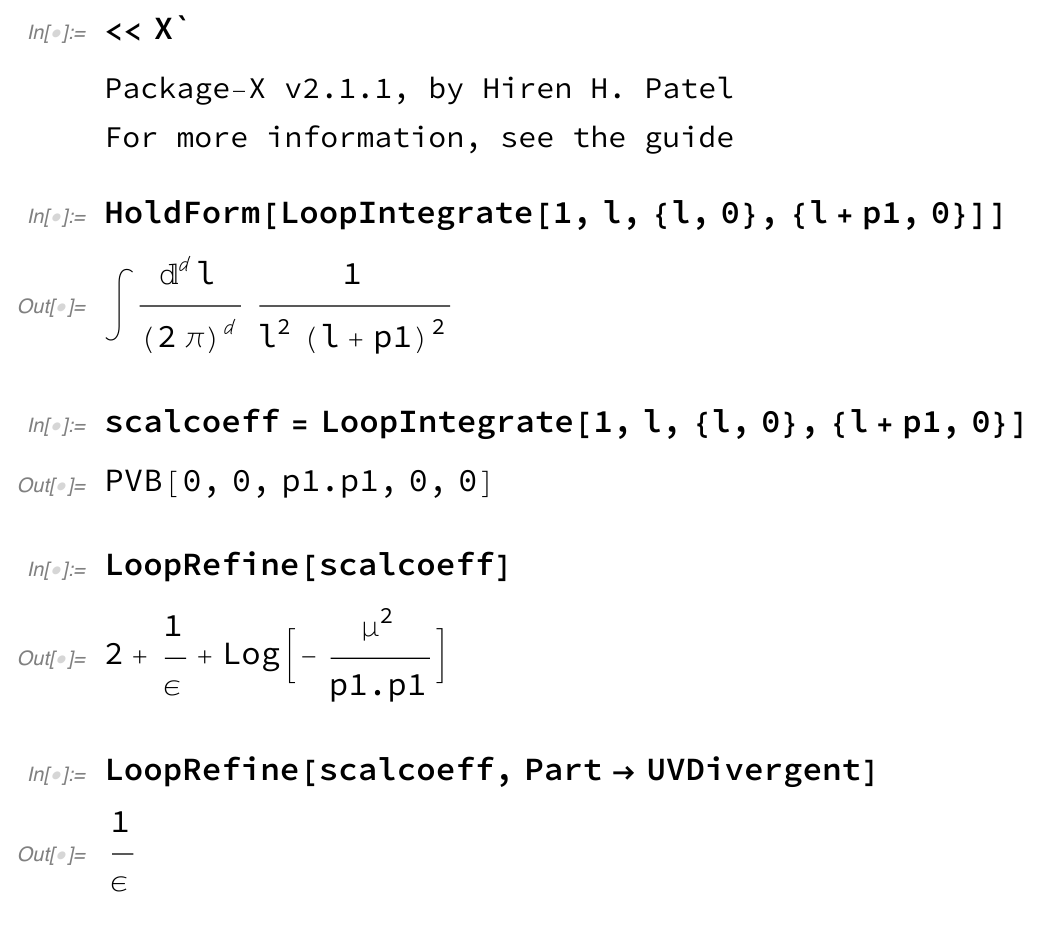}
\caption{Loading \patel in \wolfram and extracting the UV pole of the scalar two-point function}
\label{looprefine_fig}
\end{figure}

The syntax of the routines employed in the code is the following (see Fig. \ref{looprefine_fig}):
\begin{itemize}
\item \emph{LoopIntegrate} is the central routine of the package and performs the reduction of the tensor integrals. 
Its first argument is the numerator of the tensor integral, the second is the loop momentum, the following pair given as
arguments have the structure $(mom,mass)$ where $mom$ is the momentum and $m$ is the mass of the propagating particle for each propagator in the loops.
The mass is always $0$ for us. Various options can be given, for which we refer the reader to the \patel guide, included in the \wolfram interactive documentation upon
successful installation of the package. 
\item \emph{LoopRefine} converts the Passarino-Veltman coefficient functions to analytic expressions which can be readily evaluated numerically, 
         extracting the ultraviolet pole from the scalar two-point function. Various options can be given, among which is \emph{Part $\rightarrow$ UVdivergent},
         which makes the routine extract only the ultraviolet pole of the expression; for other options we refer the reader to the \patel guide.  
\end{itemize}

In the following, we discuss in detail the 1 loop counterterms and the Ward identities they are subjected to.
We refer the reader to the public repository for further details. 
This is actually not the first time that the four-point function counterterms were computed, as
they were already worked out in~\cite{Coriano:2012dg}, where their relation to the trace anomaly - to be discussed shortly - 
was exploited to explore dilaton interactions in the conformal limit of the Standard Model. 
Another application was explored in~\cite{Coriano:2013xua}, where a recursive relation allowing to compute
fully traced correlators of the energy-momentum tensor was discovered. \\

The term we add to the action in our generating functional $\mathcal W$ in order to renormalize our 
(unrenormalizable) theory in (\ref{genfunc}) is the following,
\beq\label{CounterAction}
S_{counter} = - \frac{\mu^{-2\epsilon}}{\epsilon}\, \int d^d x \sqrt{g} \bigg( \beta_a\, F + \beta_b\, G\bigg),
\eeq
with $d = 4-2\epsilon$, containing the squared Weyl tensor $F$ and the Euler density $G$, defined in Appendix \ref{conventions}.
Since the integral of the Euler density is a topological invariant for $d=4$, it can be proven that the Euler part of the three and four-point 
counterterms derived from (\ref{CounterAction}) are actually finite, because the functional derivatives of the integral of the Euler density in $d=4-2\epsilon$ dimensions 
are $\propto \epsilon$. This means that the contribution subtracted through the term $\propto \beta_{b}$
in (\ref{CounterAction}) is finite and, thus, this amounts to a choice of the renormalization scheme.
To the best of our knowledge, this was first argued and shown to be true for $d=2$ in~\cite{Deser:1993yx}, while
a very detailed technical derivation in momentum space recently appeared in~\cite{Bzowski:2017poo}, to which we refer the interested reader, 
particularly for the case of the Euler counterterm of the three-point EMT correlation function in $d=4$. 
The latter is studied in great detail on the grounds of an elegant form factor decomposition of the three-point correlator,
which unveils the hidden dimension-dependent degeneracy underpinning the vanishing of the functional variations of the integrated Euler density in $4$ dimensions. 
On our side, since we do not have such a general decomposition of our four-point function at our disposal yet,
we do not attempt the generalization of this procedure for the four-point function, but we employ another approach, discussed in Appendix A.2 of~\cite{Bzowski:2013sza}, 
to explicitly show the vanishing of the Euler counterterm for both the three and four-point functions in the notebook \emph{vanishing\_euler\_ct.nb}, 
discussing the procedure in Appendix \ref{eulerisnull}. This explicitly proves that, also for the four-point function, the Euler contribution to the counterterm
serves as a finite renormalization which has the ultimate purpose of yielding a correlator whose trace anomaly has the expected form (\ref{Anomalies234}). \\

The counterterm (\ref{CounterAction}) could be further supplemented by additional and explicitly finite terms $\propto \beta_{a}$, 
but we chose not to include such a discussion, which can be found in many papers on EMT correlators~\cite{Osborn:1993cr,Coriano:2012wp,Serino:2014ysa}.

For the general $nT$ correlator, the counterterm action (\ref{CounterAction}) generates the $n$-point vertex 
\beq
-\frac{\mu^{-2\epsilon}}{\epsilon}\, \bigg(
\beta_a\, D_F^{\muu\nuu\dots\mu_n\nu_n}(x_1,\dots,x_n) + 
\beta_b\, D_G^{\muu\nuu\dots\mu_n\nu_n}(x_1,\dots,x_n)\bigg) \, ,
\eeq
where
\bea
D_F^{\muu\nuu\dots\mu_n\nu_n}(\xu,\dots,x_n)
&=&
2^n \, \frac{\delta^n}{\delta g_{\muu\nuu}(x_1)\dots g_{\mu_n\nu_n}(x_n)}
\int\,d^d x\,\sqrt{g}\, F\, ,
\label{DF}\\
D_G^{\muu\nuu\dots\mu_n\nu_n}(\xu,\dots,x_n)
&=&
2^n \, \frac{\delta^n}{\delta g_{\muu\nuu}(x_1)\dots g_{\mu_n\nu_n}(x_n)}
\int\,d^d x\,\sqrt{g}\, G\, .
\label{DG}
\eea
The momentum space couterterms are defined via the same Fourier transform defining the momentum space correlation functions (\ref{fourier}). \\

We have derived in detail the first functional variation of the integral of a general expression which is quadratic in the Riemann tensor in
a shared appendix of~\cite{Coriano:2012wp,Serino:2014ysa}. 
Here we provide only the final result.  \\

Let $K = \big(a\,R^{\alpha\beta\gamma\delta}R_{\alpha\beta\gamma\delta} + b\,R^{\alpha\beta}R_{\alpha\beta} + c\, R^2 \big)$, 
with $a$, $b$ and $c$arbitrary real numbers; then 
\bea
&&
\frac{\delta}{\delta g_{\muu\nuu}(x)}\,  \int\,d^d x\,\sqrt{g}\, K\, =
\nn \\
&&
\sqrt{g}\, \bigg\{\frac{1}{2}g^{\muu\nuu}K- 2a\, R^{\mu\alpha\beta\gamma}{R^\nu}_{\alpha\beta\gamma}
+ 4a\,R^{\muu\alpha}{R^\nuu}_\alpha - (4a+2b)\, R^{\muu\alpha\nuu\beta}R_{\alpha\beta} - 2c \, R R^{\muu\nuu}
\nn\\
&& \hspace{8mm}
+ \, (4a + b)\,\Box{R^{\muu\nuu}} + (4c + b)\,g^{\muu\nuu}{R^{\alpha\beta}}_{;\alpha;\beta}
- (4a+2b+4c){{R^{\nuu\beta}}_{;\beta}}^{;\muu}\bigg\}\, .
\label{firstfunc}
\eea
From the equation above, the counterterm action (\ref{CounterAction}) and taking traces in $d$ dimensions, 
since we are working in DR, one can derive the trace anomaly observing that
\bea
g^{(d)}_{\muu\nuu}\, \frac{\delta}{\delta g_{\muu\nuu}(x)}\,  \int\,d^d x\,\sqrt{g}\, F \, &=& - \sqrt{g}\,\frac{\epsilon}{2}\, \bigg(F- \frac{2}{3}\, \square R \bigg)\, ,
\nn \\
g^{(d)}_{\muu\nuu}\, \frac{\delta}{\delta g_{\muu\nuu}(x)}\,  \int\,d^d x\,\sqrt{g}\, G \, &=& - \sqrt{g}\,\frac{\epsilon}{2}\, G \, ,
\label{dufftraces}
\eea
where we have distinguished the $d$-dimensional trace with a superscript on the metric tensor.

Using these expressions, the renormalized two, three and four-point correlators in momentum space can be written as
\bea
&& \hspace{-25mm}
<T^{\muu\nuu}(-\pdd)T^{\mud\nud}(\pdd) >_{ren} \,=\,
<T^{\muu\nuu}(-\pdd)T^{\mud\nud}(\pdd) >_{bare} -
\frac{1}{{\epsilon}}\, \beta_a\, D_F^{\muu\nuu\mud\nud}(-\pdd,\pdd) \, ,
\label{Ren2PF}
\nn \\
&& \hspace{-25mm}
<T^{\muu\nuu}(\pu)T^{\mud\nud}(\pdd)T^{\mut\nut}(\pt) >_{ren} \,=\,
<T^{\muu\nuu}(\pu)T^{\mud\nud}(\pdd)T^{\mut\nut}(\pt) >_{bare}
\nn \\
&& 
-\, \frac{1}{{\epsilon}}\,\bigg(
\beta_a\, D_F^{\muu\nuu\mud\nud\mut\nut}(\pu,\pdd,\pt) + 
\beta_b\, D_G^{\muu\nuu\mud\nud\mut\nut}(\pu,\pdd,\pt) \bigg)\, ,
\label{Ren3PF}
\nn \\
&& \hspace{-25mm}
<T^{\muu\nuu}(\pu)T^{\mud\nud}(\pdd)T^{\mut\nut}(\pt)T^{\muq\nuq}(\pq) >_{ren} \,=\,
< T^{\muu\nuu}(\pu)T^{\mud\nud}(\pdd)T^{\mut\nut}(\pt)T^{\muq\nuq}(\pq)>_{bare}
\nn \\
&&
-\, \frac{1}{{\epsilon}}\,\bigg(
\beta_a\, D_F^{\muu\nuu\mud\nud\mut\nut\muq\nuq}(\pu,\pdd,\pt,\pq) + 
\beta_b\, D_G^{\muu\nuu\mud\nud\mut\nut\muq\nuq}(\pu\pdd,\pt,\pq) \bigg)\,  .
\label{Ren4PF}
\eea
From these relations and from (\ref{transverse2t}), (\ref{transverse3t}) and (\ref{transverse4t}) it is apparent that the counterterms are 
related to each other by the same transverse Ward identities relating the EMT correlators.  
One can also separately check these identites for Weyl and Euler counterterms just by writing them down and equating 
the coefficients of $\beta_a$ and $\beta_b$. This is done in the second part of the  file \emph{functional\_derivatives.nb}. 

Anomalous Ward identities for our counterterms can be derived through up to three more functional derivatives of Eqs. (\ref{dufftraces}), 
so that identities which are completely analogous to (\ref{trace2t})-(\ref{trace4t}) emerge,
the only difference being that now traces are taken in $d$ dimensions. We report them separately for the Weyl and Euler counterterms
in the first channel: of course analogous identities -obtained by proper permutations of indices and momenta- hold in all channels. 
\bea
\delta^{(d)}_{\muu\nuu} D_F^{\muu\nuu\mud\nud}(-\pdd,\pdd)  &=& - \frac{\epsilon}{2}\, \frac{2}{3}\, \left[ \square R \right]^{\mud\nud}(\pdd) \, , 
\nn \\
\delta^{(d)}_{\muu\nuu} D_G^{\muu\nuu\mud\nud}(-\pdd,\pdd)  &=& 0\, ,
\label{trace2tcount}
\eea
\bea
\delta^{(d)}_{\muu\nuu} D_{F}^{\muu\nuu\mud\nud\mut\nut}(\pu,\pdd,\pt) 
&=& 
- D_F^{\mud\nud\mut\nut}(-\pdd,\pdd)  - D_F^{\mud\nud\mut\nut}(-\pt,\pt)
\nn \\
&& 
- \frac{\epsilon}{2} \, \bigg[ \bigg( F - \frac{2}{3}\, \sqrt{g}\, \square R\bigg) \bigg]^{\mud\nud\mut\nut}(\pdd,\pt) \, , 
\nn \\ 
\delta^{(d)}_{\muu\nuu} D_{G}^{\muu\nuu\mud\nud\mut\nut}(\pu,\pdd,\pt) 
&=& 
- D_G^{\mud\nud\mut\nut}(-\pdd,\pdd)  - D_G^{\mud\nud\mut\nut}(-\pt,\pt)
\nn \\
&&
- \frac{\epsilon}{2} \bigg[G \bigg]^{\mud\nud\mut\nut}(\pdd,\pt) \, , 
\label{trace3tcount}
\eea
\bea
\delta^{(d)}_{\muu\nuu}\, D_{F}^{\muu\nuu\mud\nud\mut\nut\muq\nuq}(\pu,\pdd,\pt,\pq) 
&=& 
-\, 2\, D_{F}^{\mud\nud\mut\nut\muq\nuq}(-\pt-\pq,\pt,\pq) 
\nn \\
&& \hspace{-30mm}
- 2\, D_{F}^{\mud\nud\mut\nut\muq\nuq}(-\pdd-\pq,\pdd,\pq)
- 2\, D_{F}^{\mud\nud\mut\nut\muq\nuq}(-\pdd-\pt,\pdd,\pt) \, , 
\nn \\
&& \hspace{-30mm}
-\, \frac{\epsilon}{2}\, \bigg(\left[\sqrt{g}\, F\right]^{\mud\nud\mut\nut\muq\nuq}(\pdd,\pt,\pq) 
-\, \frac{2}{3}\,  \left[\sqrt{g}\,\Box R \right]^{\mud\nud\mut\nut\muq\nuq}(\pdd,\pt,\pq)\bigg)
\nn \\
\delta^{(d)}_{\muu\nuu}\, D_{G}^{\muu\nuu\mud\nud\mut\nut\muq\nuq}(\pu,\pdd,\pt,\pq) 
&=& 
-\, 2\, D_{G}^{\mud\nud\mut\nut\muq\nuq}(-\pt-\pq,\pt,\pq) 
\nn \\
&& \hspace{-30mm}
- 2\, D_{G}^{\mud\nud\mut\nut\muq\nuq}(-\pdd-\pq,\pdd,\pq)
- 2\, D_{G}^{\mud\nud\mut\nut\muq\nuq}(-\pdd-\pt,\pdd,\pt)
\nn \\
&& \hspace{-30mm}
-\, \frac{\epsilon}{2}\, \left[\sqrt{g}\, G\right]^{\mud\nud\mut\nut\muq\nuq}(\pdd,\pt,\pq) \, .
\label{trace4tcount}
\eea
It is apparent that all these constraints are not trivial to satisfy.
We checked all of them for all anomalies and counterterms and, thus, ensured that these were correctly computed.

As mentioned before, the ultraviolet pole of our four-point function is found to coincide with the overall counterterm for the scalar field $4T$. \\
The reason why we perform this preliminary check is twofold.
\begin{itemize}
\item The ultraviolet pole of our four-point function is highly non trivial, since there are 16 diagrams contributing to it, so that
         getting it right is a solid first check of the diagrammatic expansion. Beside, being a polynomial in the external momenta, 
         the ultraviolet pole of our four-point function is much more manageable than the whole correlator, which requires much more memory and computing time. 
         This makes it an ideal tool to quickly scan for potential mistakes in the Feynman expansion.
\item The trace Ward identities connect the counterterms to the trace anomalies and, so, are a test for the latter as well. 
         The trace anomalies are present also in turn in the trace Ward identities for the correlation functions (\ref{trace2t})-(\ref{trace4t}), 
         which we must test in order to ensure the correctness of the four-point function. Thus, having them tested in advance reassures us
         about the correctness of the trace identities, so that any mismatch emerging from them can be quite surely traced back to the four-point function. 
\end{itemize}
One last comment is in order. The Euler characteristic -the integral over all space of the Euler density- is topologically invariant, i.e. it does not depend on the specific 
metric tensor used in the integrand, so its derivatives w.r.t. the metric vanish identically for $d=4$, meaning they are $\propto \epsilon$ in $d=4-2\,\epsilon$, 
as already discussed below Eq. (\ref{CounterAction}). Thus, checking that its functional derivatives correctly contribute to our counterterms 
because they remove terms $\propto 1/\epsilon$ in our regularized correlator effectively amounts 
to testing that two finite sets of terms, one in the regularized correlator and the other in the counterterm, exactly match. 
Though our computer algebra stays ``unaware'' of the finiteness of these contributions, since the proportionality to $\epsilon$ is hidden, 
this is clearly a sensible test. \\

Once all these preliminary tests have been successful, we can be reasonably confident about the diagrammatic expansion. 
What is left it the computation of the whole four-point function and the test of the transverse and trace Ward identities.
Plenty of details are available in the files \emph{correlators\_calculation.nb} and \emph{ward\_identities.nb}.

The purpose of the following section is to go into the necessary technical details about the calculation. 

\section{Into the full calculation}\label{mathematica}

It is time to illustrate in detail how we employ our tools to nail down the whole four-point function. 
In the first part of this section we review the Feynman diagram computation, in the next two we give a survey of the checks of the Ward identities.
A few snippets of code are provided in this section, but we encourage the interested reader to explore the notebooks, 
which are documented in great and hopefully sufficient detail.   

\subsection{The calculation of the four-point correlator}

The calculation is performed by the file \emph{correlators\_calculation.nb}.
The code employs \patel to automate the Passarino-Veltman reduction of the many tensor integrals one encounters.

The notebook starts by loading the vertices and counterterms stored in the \emph{all\_functional\_derivatives} file, 
which must be generated beforehand by running \emph{functional\_derivatives.nb}. 
After that, the notebook is divided into three parts, one for each of the three computed correlators.
The structure of these three parts is similar and unfolds as follows.
\begin{itemize}
\item The numerators of the contributing Feynman diagrams are constructed. As detailed in section \ref{setup}, there is only 1 for the $2T$, 
         while there are 4 for the $3T$ and 16 for the $4T$. An image of each computed diagram is loaded in the notebook right before the line computing its numerator.
         In the case of the two-point function, the known result for general values of the improvement parameter (see Eq. \ref{scalarAction})
         is presented and rederived for the reader's convenience. The case with $\chi = 1/6$ corresponds to the conformally invariant case and is eventually selected.\footnote{See also the paragraph
         just before section 2.1 about the choice $\chi =1/6$ vs the general case $\chi = 1/4 (d-2)/(d-1)$}
\item The \emph{LoopIntegrate} routine is invoked to perform the tensor reduction of the diagrams.
         For the three bubbles in the three-point function and for all the diagrams contributing to the four-point function, only blueprint diagrams with generic momenta are computed.  
         The actual diagrams are obtained by replacing the generic momenta with the sets classified in section \ref{setup}. 
         This is topical for the four-point function, due to the memory it requires (see next point). 
\item The \emph{LoopRefine} routine is employed to extract the ultraviolet poles from each diagram and to sum them up in order to identify the ultraviolet divergence of the correlator.
         This is then compared to the corresponding counterterm and exact matching is found, so the whole correlators are stored in the files 
         \emph{T2\_scalar}, \emph{T3\_scalar} and \emph{T4\_scalar}. The first two contain just the full two and three-point correlators, whereas \emph{T4\_scalar} stores only the four blueprint diagrams, 
         because of the amount of memory the full result would otherwise require: the blueprint for the square diagram takes roughly $2.3$ GB of memory by itself. 
         This is before substituting an explicit set of momenta, which makes it even lengthier because some of them are the sum of more momenta and before exploiting 
         the momentum conservation constraint $\pu \rightarrow -(\pdd+\pt+\pq)$. 
\end{itemize}

The results presented in this section were relatively straightforward to get, although one of them requires quite some time:
the extraction of the ultraviolet pole of the square diagram contributing to the four-point function takes roughly $20$ mins even when 
parallelized among 8 kernels, each one running on a different core of a 2,4 GHz Intel Core i9 CPU.
Nevertheless, despite the time required, the amount of memory used to accomplish any of the tasks in this notebook never exceeds 
$4$ GB, which makes it easy to execute on any modern laptop.

Unfortunately, this is not the case for the tests of the Ward identities for the $4T$ correlator, as we will explain in a while.
The next two sections deal with the content of the file \emph{ward\_identities.nb}.

\subsection{Analytical checks of the Ward identities for the $2T$ and $3T$}

\begin{figure}[h!]
\includegraphics[scale=1.0]{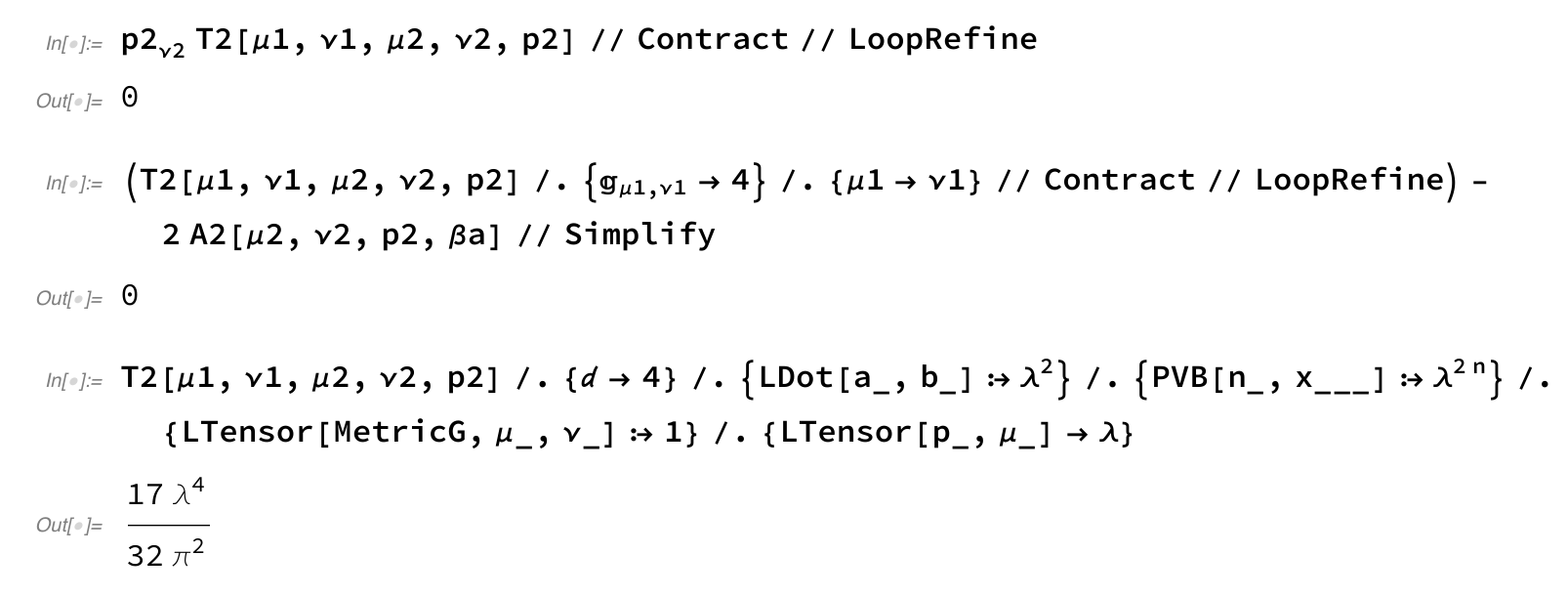}
\caption{Check of the transverse and trace Ward identities as well as of the naive scaling dimension for the two-point function.}
\label{snippet_ward_2t_fig}
\end{figure}

Checking the transverse and trace Ward identities for the two-point function, Eqs. (\ref{transverse2t}) and (\ref{trace2t}), is trivial.
The first two lines in fig. \ref{snippet_ward_2t_fig} should be self explanatory, once the reader has understood the essential purpose 
of the \emph{LoopRefine} routine explained in section \ref{counterterm} and that the $\bold{A2}$ object is just the anomaly on the rhs of (\ref{trace2t}).

It is time to explain how to check the naive scaling dimension of the correlators. The underpinning idea is very simple. 
This dimension is called naive because proper scaling accounts for the effect of logarithms dependent on the ultraviolet scale $\propto \log(p^2/\mu^2)$, 
which come from the two-point scalar integrals. 
Since every term in our EMT correlators must be of dimension $(momentum)^4$, we can replace it with its dimension, 
which we choose to denote with $\lambda$ in the code. So, for instance, $p_{\mu} \rightarrow \lambda, g_{\mu\nu} \rightarrow \lambda^{0}$. 
Once all the replacements have gone through, we expect to get just a real number times $\lambda^4$, 
which is what is shown in the last input line of the snippet below and what is found for the higher-point functions as well.
 \\
One might ask why we do not undertake a check of the full dilatation identity, which would be, in $d=4$, something like
\beq\label{dilatation}
\bigg(4 - \sum_{i=1}^n p^{\lambda}_i \frac{\pd}{\pd p^{\lambda}_i}\bigg) <T^{\muu\nuu}(\pu)\dots T^{\mu_n,\nu_n}(p_n) > = \text{anomalous terms}\, .
\eeq
The reason is that checking the trace identities (\ref{trace2t})-(\ref{trace4t}) already implies satisfying the scale invariance requirement.
Indeed, the trace anomaly equation (\ref{traceanomaly}) is obtained by studying the transformation of the generating functional of a field theory embedded 
in curved space under a local rescaling of the metric tensor, which is called a Weyl transformation. 
This is a more general transformation than a global rescaling of coordinates and, for Lagrangian field theories which are 
at most quadratic in their fields derivatives, is effectively equivalent to full-fledged conformal invariance~\cite{Iorio:1996ad}.
This means in turn that, since this symmetry is broken only by the trace anomaly, which can be seen as a by-product of the ultraviolet renormalization 
of the EMT,  it follows that, if an EMT correlator satisfies its anomalous trace identity, the correlator must also be scale invariant in the sense of (\ref{dilatation}). 
In particular this implies that, if its momenta are uniformly rescaled by a global factor $p \rightarrow \lambda p$, it
must change by an overall factor $\lambda^4$ plus additional contributions due to the logarithms associated to the regularization scale $\mu^2$,
introduced to consistently regularize ultraviolet divergencies $\propto \log(p^2/\mu^2 ) \rightarrow \log(p^2/\mu^2 ) + \log \lambda^2$.
These are the contributions that, acted upon by the differential operators in Eq. (\ref{dilatation}), would render the anomalous term on the rhs. 
The fact that in Fig. \ref{snippet_ward_2t_fig} we are just cheking the naive scale dimension is due to the replacement 
of the two-point scalar integrals with $\lambda^0$, which does not take into account the logarithmic contribution.

Now, the reason why we are discussing the naive scale identity is that its check can be done analytically also for the complicated four-point correlator 
and, as such, will prove very useful in the last part of this section, when numerical stability of our results is discussed. \\

Next we come to the tests of the Ward identities for the three-point function, which is more involved but was 
already done in~\cite{Coriano:2012wp} in pretty much the same way, which we reproduce here.
The idea behind the procedure for both transverse and trace identities is the same and quite simple.
\begin{itemize}
\item Load the tensor basis which spans the space of rank-5 or rank-4 tensors dependent on 2 independent momenta to which 
         belongs the correlator contracted with the momentum in (\ref{transverse3t}) or with the metric tensor in (\ref{trace3t}); 
         assign to them the correct indices (those which survive after contraction).
\item Build the lhs and rhs for each Ward identity separately.
\item Isolate the coefficients of each tensor in the basis for both the lhs and the rhs, subtract them from each other and check that for all of them the result is zero. 
\end{itemize}
We have checked all three transverse identities (\ref{transverse3t}) and trace identities  (\ref{trace3t}). 
For further details, the reader is encouraged to look at the code, where she will also encounter the \emph{KallenExpand} routine, 
which plugs the explicit expression for the K\"all\'en polynomial $\lambda(a,b,c) = a^2+b^2+c^2-2ab-2ac-2bc$
in place of its symbolic representation.

\subsection{Numerical checks of the Ward identities for the $4T$ correlator}

This is the most slippery and time consuming part of our project, because the whole $4T$ correlator
is just too big to be dealt with analytically within reasonble time with an ordinary computer, even one with a $64$ GB RAM like the one at our disposal.
As we mentioned above, the blueprint of the square diagrams contributing to the $4T$ are the culprits,
each one of them occupying alone almost $4$ GB of memory space after the proper momenta assignments. 
These assigments are such that the second argument of each diagram is the sum of two external momenta (see the notebook and the diagrams 
of Section \ref{setup}) which, considering the amount of terms it consists of, significantly  increases the necessary memory. 
The first feature which is required from the needed computer is quite some RAM: at least 40 available GB, in order for the computation not to crush.
The second crucial feature is to have more than one available CPU core to perform calculations, in order to reduce the required time to a reasonable 
window through parallelization.
In our case, we had $8$ kernels and a $64$ GB RAM at our disposal on our machine, which allowed us to complete the test of all the 9 Ward identities for the $4T$ correlator 
in slightly less than $1$ hour.\footnote{We are perfectly aware that state-of-the-art numerical calculations, e.g. for multiloop LHC phenomenology, 
require tens of thousands of CPU hours to complete, dwarfing our case. The spirit of our endeavor is different though: we are ultimately after some
clever way to handle these results on even less powerful devices than ours in a near future.}

In order to speed up our numerical effort, we resorted to the \denner package, which connects \patel with the COLLIER library (written in C language) in order to
provide fast numerical evaluation of the scalar integrals. A fundamental component of the numerical evaluation procedure is to isolate the ultraviolet poles
from the rest of the correlators.

The steps of the testing procedure explained for the $3T$ Ward identities hold here as well, with two caveats. First, the tensor bases to be used are now much bigger, due to
the increased number of indices from $5$ to $7$ for transverse Ward identities and from $4$ to $6$ for trace Ward identities, beside the fact that we have one more independent momentum
in our correlator w.r.t. the three-point case; second, the matching of the coefficients must now be checked numerically in order to guarantee sufficient speed
and, for most common computers, not to crush. 

On the ground of what was explained so far, one might think that subtracting the counterterm for the $4T$ after applying the \emph{LoopRefine} 
routine to the computed correlator -even parallelizing it- would do the job: it would indeed, but that would be much too slow. 

What one can do, instead, is to resort to \denner, which can perform direct numerical evaluation of the Passarino-Veltman coefficient functions bypassing the analytic expansion.
The task is parallelized in the notebook through the \emph{ParallelMap} \wolfram routine, which distributes the task among the various available kernels.
The parallelized task is in turn a sequence of two steps.
\begin{itemize} 
\item The \emph{SeparateUV} routine of \denner isolates the UV pole in each coefficient function.
\item The \emph{numrep} set of rules, defined above in the notebook, replaces all the scalar products of momenta with the corresponding numerical values 
         obtained after choosing a completely off-shell configuration which respects $\pu+\pdd+\pt+\pq = 0$. The input of numerical values into the scalar integrals
         automatically triggers \denner to invoke the COLLIER library and to evaluate the scalar coefficients numerically. 
         We must specify a few things in order to have \denner run as we need, such as the accuracy required in the evaluation of scalar integrals, 
         which we set to $10^{-10}$ \footnote{This is $2$ orders of magnitude smaller than the precision required for the numerical test, $10^{-8}$}, 
         and the maximum tensor rank of our integrals, $8$. The ultraviolet regularization scale $\mu$ is set by default to $1$, but this value can be changed (see the next part of this section). 
\end{itemize}
This step is the real bottleneck of the whole procedure, for it takes $\simeq 40$ minutes to complete for most of the momenta configurations we have tested, which are
\begin{eqnarray}
\left[\pu,\pdd,\pt,\pq \right] = \,
&&
\left\{ \begin{array}{c}
\hspace{15mm} \left[(5,-3,0,3),(-1,1,0,1),(-2,1,0,-1),(-2,1,0,-1)\right] \\
\hspace{20mm} \left[(-2,-3,9,1),(1,2,4,1),(3,-6,-2,0),(-2,7,-11,-2)\right] \\
\hspace{24mm} \left[(1,1,1,1),(1,-1,-1,-1),(-1,\frac{1}{2},\frac{1}{2},\frac{1}{2}),(-1,-\frac{1}{2},-\frac{1}{2},-\frac{1}{2})\right] \\
\bigg(\frac{1}{\sqrt2}-\frac{1}{\sqrt{3}}\bigg)\times \left[(1,1,1,1),(1,-1,-1,-1),(-1,\frac{1}{2},\frac{1}{2},\frac{1}{2}),(-1,-\frac{1}{2},-\frac{1}{2},-\frac{1}{2})\right] \\
\end{array} \right. \quad\quad
\label{momentalists}
\end{eqnarray}
Please notice that the fourth configuration is given by the momenta in the third one rescaled by the prefactor in round brackets.
This will be topical for the upcoming discussion of the numerical precision and stability of the tests.

\begin{figure}[h!]
\includegraphics[scale=1.0]{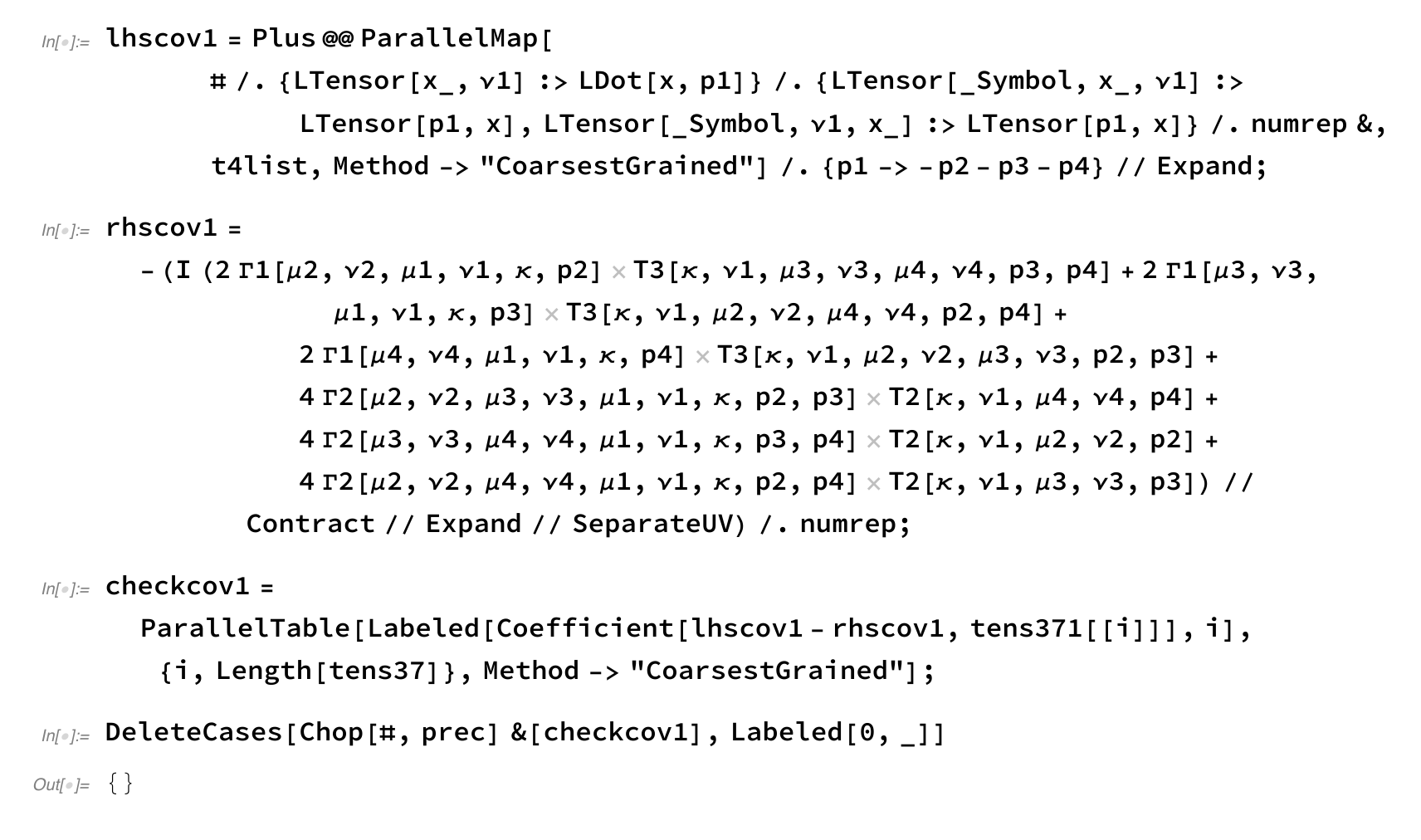}
\caption{Parallelized numerical comparison of the scalar coefficients over all the tensors in the base \emph{tens371} (3 momenta, rank 7, indices assignments for the 1st identity) 
of the lhs and rhs of the first transverse Ward identity in (\ref{transverse4t}). If smaller than the threshold \emph{prec}$=10^{-8}$ parameter, 
the result is subsequently chopped out of the list. The final result is an empty list, so the test is successful.}
\label{snippet_transverse_ward_4t_fig}
\end{figure}

Once the numerical evaluation of the scalar coefficients has been performed, what is left in order to test the Ward identities is to contract
the listed diagrams with either one of the momenta or the metric tensor, perform the few more numerical replacements on the scalar products
this produces, perform the same operations on the corresponding rhs (see Eqs. \ref{transverse4t} and \ref{trace4t}) and compare the 
scalar coefficients one by one. Then, if all the differences are acceptably small, we can declare the test successful.
The corresponding snippet of code is given in Fig. \ref{snippet_transverse_ward_4t_fig}, which we comment below
\begin{itemize}
\item \emph{lhscov1} is the lhs of (\ref{transverse4t}) and the set of rules in the curly brackets are the (faster) equivalent of the contraction with the $\pu$ momentum.
         When momentum conservation is employed at the end, it can affect only tensor structures, as it must in order not to miss any terms, since the tensor basis \emph{tens371}
         depends only on 3 independent momenta.
\item \emph{rhscov1} is the rhs of Eq. (\ref{transverse4t}), the $\Gamma_1$ and $\Gamma_2$ symbols being defined in Appendix \ref{Vertices},
whereas the $T2$ and the $T3$ functions are obviously the two and three-point correlators. 
\item \emph{checkcov1} is a table, evaluated in parallel, made up by the differences of all scalar coefficients of the lhs and the rhs over the basis of rank 7 tensors with 3 independent momenta. 
         The additional function \emph{Labeled} just marks every difference with the number of the tensor in the list, which could be needed after an unsuccessful test to track down 
         the tensors whose coefficients on the lhs and rhs of the identity do not match.
\item The final line simply removes from the list all those numbers which were set equal to $0$ through the \emph{Chop} routine, because both their real 
         and imaginary parts were under the \emph{prec} threshold, which we set to $10^{-8}$ for the first three momentum configurations listed in (\ref{momentalists}).
         A dedicated discussion of the fourth momentum configuration, for which the precision threshold we manage to reach is not even close, being just $10^{-1}$, aims at proving that
         this is definitely an expected issue of numerical instability. To this end, the fact that we can easily check analytically  naive scaling dimension is useful.  
\end{itemize}
We will not delve any further into the discussion of the tests of the other Ward identities, as all of them are pretty much the same, modulo indices and momenta reshuffling.


\subsection{Discussion about numerical stability}

The last point we wish to discuss before coming to our conclusions is the issue of numerical stability, which is raised when one tries to check
the four-point functions Ward identities for momenta configurations like the last one in (\ref{momentalists}), which was purportedly
chosen to be a rescaled version of the back-to-back third configuration, for which the test is precise through $8$ decimal figures, 
just as for the former $3$ as well.

This is when the test of the naive scale invariance of our correlators comes in handy, as we finally explain, reassuring us that the problem is only due to numerical instability.
Now, the naive scaling dimension of every EMT correlator is $(momentum)^4$, as one can see analytically for all correlators.
This means that both the lhs and the rhs of every identity in (\ref{transverse4t}) and (\ref{trace4t}) should naively scale by the same factor $\lambda$.
To be completely precise, one must not forget to take into account the logarithmic contributions
associated with scalar two-point functions which, as mentioned in section \ref{ward}, behave under rescaling like
$\log(p^2/\mu^2) \rightarrow \log(\lambda^2 p^2/\mu^2) = \log(p^2/\mu^2) + \log \lambda^2$.

\begin{figure}[h!]
\includegraphics[scale=1.0]{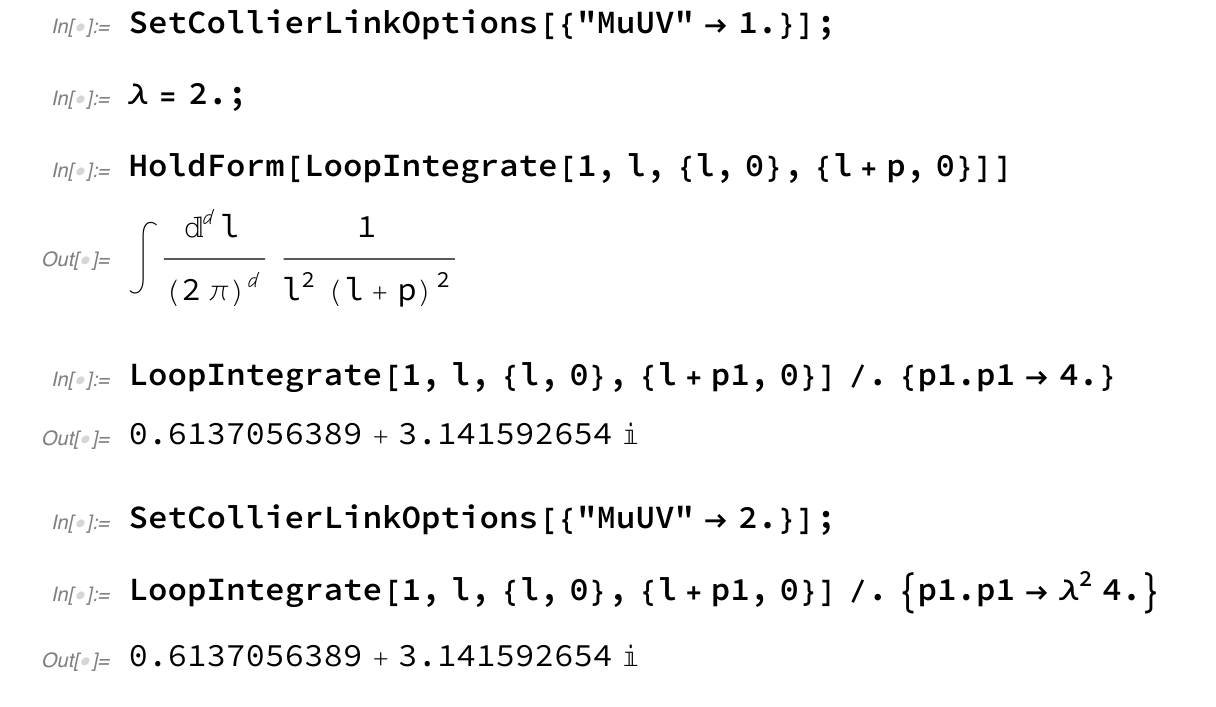}
\caption{The scaling behaviour of the two-point function: scaling the regularization scale by the same factor as the momentum makes 
it behave as if its scaling dimension were the naive one, i.e. $0$.}
\label{snippet_b0scaling_fig}
\end{figure}

If it comes to numerically checking the rescaled identities, we can take care of this problem in two ways: 
\begin{itemize} 
\item by rescaling the regularization scale as well by $\mu \rightarrow \lambda \mu$, using the dedicated initialization option of the \denner package:
the snippet of code in Fig. \ref{snippet_b0scaling_fig} illustrates the invariance of the two-point function after rescaling both the momentum and the regularization scale.
\item by leaving everything as it is, since the same two-point functions are present on both the lhs and rhs of the Ward identity, so that extra terms should coincide as well. 
\end{itemize}
This proves that, if the identities are satisfied for a given momentum configuration, they should be identically satisfied for the rescaled configuration as well. 
Testing both possibilities is useful to realize that the extra logs coming from the rescaling of the momenta are not the source of the numerical instability. 
Indeed, what we find for our fourth momentum configuration, whether we take care of $\mu$ in the first way suggested above or not, is that the precision up to which the
Ward identity is numerically satisfied does not exceed $10^{-1}$ for some scalar coefficients in the tensor bases.

Nevertheless, since we proved that this result must vanish exactly too, the fact that the numerical agreement is excellent for such diverse momenta 
configurations as the first three leads us to the conclusion that our calculation is correct,
although its numerical evaluation suffers from numerical instability, which is nevertheless to be expected for such huge expressions. 

Of course, one can think of ways to improve our numerical tests.

The naive way could be to pursue a brute force approach: one would feed the numerical routines higher and higher precision inputs 
(\wolfram can numerically evaluatae exact numbers to arbitrary numerical precision), but this would take an even heavier toll on the memory of the machine,
which is already quite stretched with the required accuracy standards. A batch of several machines on which we could parallelize our numerical tests could do the job, of course. 
We do not have access to an integrated \wolfram deployment of this kind, but we firmly believe that the discussion above has clarified 
that this would add neither any further understanding nor improvement to our result, not least because, once one has a much higher available computing power, 
the identities could be tested analytically right off the bat.

A second way to go could be to compile our output into optimized \emph{C++} or \emph{Fortran} code. 
This is presumably feasible, but it goes behind the scope of the present work. 

These problems will stay, of course, only until a more clever way to compute the four-point function correlator is devised, which can yield a more compact result 
cutting off all the redundancies our method necessarily implies, but for the time being we consider ourselves satisfied.

\section{Conclusions and perspectives}\label{ending}%

We have computed the four-point correlation function of the energy-momentum tensor in momentum space and provided a set of 
tools which allow the interested reader to reproduce the perturbative computation of the two, three and four-point correlation functions, 
together with the explicit and detailed construction of counterterms and anomalies and the test of the transverse and trace Ward identities.
The results for the correlators are fully analytical, expressed in terms of Passarino-Veltman coefficient functions which can 
be easily either fully reduced to scalar integrals or evaluated numerically to arbitrary precision. \\
Considering the sheer dimension of our results, the main purpose of this work should be understood as
the delivery of a benchmark for numerical checks of more compact results to be obtained in the future with different strategies,
most probably the conformal bootstrap for fields with spin.

It would also be interesting to perform the same calculation for other Lagrangian conformal field theories, such as a fermion or a gauge field: 
actually, we already did this, but decided not to publish these results because they are even more massive than the scalar case 
and checking the Ward identities for the corresponding four-point correlation functions is computationally even more demanding. 
Furthermore, unlikely the case of the three-point function, the results for the scalar, fermion and gauge fields do not allow to account for the full set of constants 
which would suffice to reconstruct any four-point correlator, so one does not gain much further insight into the structure of the CFT by computing them.
Anyway, should we make progress on speeding up such checks, we will certainly update our repository.

Given the result by Dymarsky~\cite{Dymarsky:2013wla}, it would be also very useful to identify the 22 independent structures 
making up the $4T$ correlation function in 4d in momentum space, perhaps after successfully accomplishing the task in 3d, where the number of independent structures is just $5$. 
Once and if this task is accomplished, it would also be interesting to check the expected one-to-one correspondence between the momentum space 
result and the coordinate space results of~\cite{Chalmers:2000vq,Didenko:2012tv,Didenko:2013bj,Sleight:2016dba,Bonezzi:2017vha}. 
The mapping between correlators in coordinate and momentum space has been investigated in some detail in~\cite{Coriano:2012wp,Serino:2014ysa}
and shown to lead to some non trivial consistency conditions, which make this effort worth a separate work. 

Also checking conformal Ward identities for our four-point correlator in momentum space would be interesting,
but the task requires the implementation of second order differential operators and their application to a very complicated object. 
Since the Ward identities for the four-point correlator we did test were manageable only numerically, which was already a highly non trivial task, 
we did not try to figure out a way to perform the same numerical task as efficiently when working with differential operators, 
not least because our perturbative results for the lower point functions were shown in~\cite{Coriano:2018bsy} to coincide with the non perturbative
results of~\cite{Bzowski:2013sza,Bzowski:2017poo}. The latter were obtained by solving the full set of conformal Ward identities, 
so the work of implementing differential operators in our codes for the three-point function without being able to extend the procedure 
to the four-point function would have added no original output to our effort. Again, should we make progress on this point, we will update our repository.


{\bf Acknowledgments}

First, I would like to sincerely thank Krzysztof Kutak for his support throughout the development of this project.
I am very grateful to Hiren Patel for sharing insightful suggestions about \patel, to Andreas van Hameren for pointing it to my attention
and for making several useful comments about the numerical tests. Special thanks go to Misha Lublinsky for allowing me to deploy this computation 
on a powerful computer of the Ben Gurion University of the Negev. 
Useful discussions with Claudio Corian\`{o} and Matteo Maria Maglio are also gratefully acknowledged.

\appendix

\section{Conventions}\label{conventions}

\subsection{Conventions for signs and momentum space correlators definition}

The definition of the Fourier transform of the correlation function of $n$ EMT's, which holds for any other $n$-point 
coordinate space function as well, is given by
\beq\label{fourier}
\int \, d^d\xu\, \dots d^d x_n\, \left\langle T^{\muu\nuu}(\xu)\dots T^{\mu_n\nu_n}(x_n)\right\rangle \,
e^{-i(\pu\cdot \xu + \dots + p_n \cdot x_n)} = (2\pi)^d\,
\delta^{(d)}\left( \sum_{i=1}^n k_i \right)\,\left\langle T^{\muu\nuu}(\pu)\dots T^{\mu_n\nu_n}(p_n)\right\rangle \, ,
\eeq
where all the momenta are conventionally taken to be incoming. \\

The covariant derivatives of a contravariant vector $A^\mu$ and of a covariant one $B_\mu$ are respectively
\bea
\nabla_{\nu} A^\mu \equiv \pd_\nu A^\mu + \Gamma^\mu_{\nu\r}A^\r\, ,\\
\nabla_{\nu} B_\mu \equiv \pd_\nu B_\mu - \Gamma^\r_{\nu\mu}B_\r\, ,
\eea
with the Christoffel symbols defined as
\beq\label{Ch1Christoffel}
\Gamma^{\a}_{\b\g} = \frac{1}{2}g^{\a\k}\left[-\pd_\k g_{\b\g} + \pd_\b g_{\k\g} + \pd_\g g_{\k\b} \right]\, .
\eeq
Our definition of the Riemann tensor is
\bea \label{Ch1Tensors}
{R^\lambda}_{\mu\kappa\nu}
&=&
\pd_\nu \Gamma^\lambda_{\mu\kappa} - \pd_\kappa \Gamma^\lambda_{\mu\nu}
+ \Gamma^\lambda_{\nu\eta}\Gamma^\eta_{\mu\kappa} - \Gamma^\lambda_{\kappa\eta}\Gamma^\eta_{\mu\nu}.
\eea
The Ricci tensor is defined by the contraction $R_{\mu\nu} = {R^{\lambda}}_{\mu\lambda\nu}$ 
and the scalar curvature by $R = g^{\mu\nu}R_{\mu\nu}$.\\

The functional variations with respect to the metric tensor are computed using the relations
\bea\label{Ch1Tricks}
\delta \sqrt{g} = -\frac{1}{2} \sqrt{g}\, g_{\a\b}\,\delta g^{\a \b}\quad &&
\delta \sqrt{g} = \frac{1}{2} \sqrt{g}\, g^{\a\b}\,\delta g_{\a \b}  \nonumber \\
\delta g_{\mu\nu} = - g_{\mu\a} g_{\nu\b}\, \delta g^{\a\b} \quad&&
\delta g^{\mu\nu} = - g^{\mu\a} g^{\nu\b}\, \delta g_{\a\b}\, .
\eea
The variations of the Christoffel symbols w.r.t. variations of the metric tensor are tensors themselves and their expression is
\bea \label{Ch3deltaChristoffel}
\delta \Gamma^\alpha_{\beta\gamma}
&=&
\frac{1}{2}\,g^{\alpha\lambda}\big[ 
- \nabla_{\lambda}(\delta g_{\beta\gamma}) + \nabla_{\gamma}(\delta g_{\beta\lambda}) + \nabla_{\beta}(\delta g_{
\gamma\lambda})
\big]\, ,
\nn\\
\nabla_\rho \delta\Gamma^\alpha_{\beta\gamma}
&=&
\frac{1}{2}\,g^{\alpha\lambda}\big[ - 
\nabla_{\rho}\nabla_{\lambda}(\delta g_{\beta\gamma}) + 
\nabla_{\rho}\nabla_{\gamma} (\delta g_{\beta\lambda}) 
+ \nabla_{\rho}\nabla_{\beta}(\delta g_{\gamma\lambda}) \big]\, .
\eea

\subsection{Weyl invariant and Euler density in 4d }\label{Geometrical}

It is well known that the object one has to deal with in order to construct Weyl-invariant objects for general dimensions $d$
is the traceless part of the Riemann tensor, called the Weyl tensor, defined by
\beq \label{WeyldDef}
C_{\alpha\beta\gamma\delta} = R_{\alpha\beta\gamma\delta} -
\frac{1}{d-2}( g_{\alpha\gamma} \, R_{\delta\beta} + g_{\alpha\delta} \, R_{\gamma\beta}
- g_{\beta\gamma} \, R_{\delta\alpha} - g_{\beta\delta} \, R_{\gamma\alpha} ) +
\frac{1}{(d-1)(d-2)} \, ( g_{\alpha\gamma} \, g_{\delta\beta} + g_{\alpha\delta} \, g_{\gamma\beta}) R\, .
\eeq
This object enjoys the same symmetry properties of the Riemann tensor, i.e.
\beq
C_{\alpha\beta\gamma\delta} = - C_{\b\a\g\d}=  C_{\b\a\d\g} = C_{\d\g\b\a} \, ,
\eeq
and, moreover, is traceless with respect to any couple of its indices. 
It is invariant under Weyl scalings of the metric
\beq
\delta_W {C^{\a}}_{\b\g\d} = 0 \, .
\eeq

In $4$ dimensions, the only quantity which is Weyl invariant when multiplied by $\sqrt{g}$, is the Weyl tensor squared, which is
\beq \label{Ch1Weyl}
F \equiv
C^{\alpha\beta\gamma\delta}C_{\alpha\beta\gamma\delta} =
R^{\alpha\beta\gamma\delta}R_{\alpha\beta\gamma\delta} - 2\, R^{\alpha\beta}R_{\alpha\beta} + \frac{1}{3}R^2 \, .
\eeq

The other quantity with which one can construct an integral which is Weyl invariant (in fact, a constant) for general even dimensions
$d=2\,k$ is the Euler density, defined as
\beq \label{EulerdDef}
E_{2k} = \frac{1}{2^k}\, \delta_{\mu_1 a_1\nu_1b_1\dots\mu_k a_k\nu_k b_k}\, 
R^{\mu_1\nu_1\lambda_1 \kappa_1}\dots R^{\mu_k \nu_k a_k b_k}\, .
\eeq
The antisymmetric Kronecker symbol is given by
\beq
\delta_{\nu_1 a_1 \nu_2 a_2\dots \nu_n a_n } = 
n!\, \sum_{\mathcal P(a_1,\dots,a_n)}(-1)^{T_{\mathcal P}}\,  
g_{\nu_1 \mathcal P(a_1)}\dots g_{\nu_n \mathcal P(a_n)}\, ,
\eeq
where $T_{\mathcal P}$ is the number of inversions in the permutation $\mathcal P$ of the $n$ numbers $a_1, \dots a_n$.

By applying the general definition \ref{EulerdDef}, we find that in $4$ dimensions we have
\beq
E_4 \equiv G =R^{\alpha\beta\gamma\delta}R_{\alpha\beta\gamma\delta} - 4\,R^{\alpha\beta}R_{\alpha\beta} + R^2\, , \nn \\
\eeq
%

\subsection{Vanishing of the Euler counterterms}\label{eulerisnull}

In this section, we outline the proof of the claim made at the end of section \ref{counterterm}, that the Euler counterterm vanishes in $d=4$,
as all the derivatives of the integrated Euler density, which is a topological invariant. Explicit calculations can be found in the notebook \emph{vanishing\_euler\_ct.nb}. 
The derivation is inspired by the discussion of the three point function in Appendix A.2 of~\cite{Bzowski:2013sza}. 

In order to do so, we exploit the fact that in $d$ dimensions only $d$ momenta can be independent. 
If we have $d$ such momenta available, then the metric is not an independent tensor and we can rewrite it as
\beq
\delta^{\mu\nu} = \sum_{j,k = 1}^{d} p^\mu_j\, p^{\nu}_k\, \left( Z^{-1}\right)_{kj} \, , 
\label{metricton}
\eeq
where $Z$ is the Gram matrix, i.e. $Z_{ij} = (p_{i} \cdot p_{j})_{k,j=1}^{d}$.
Since, except for specific kinematic configurations, a general $n$-point function depends on $n-1$ independent momenta, because of momentum conservation,
if $n=d$, we can construct the $n$-th independent momentum using the completely antisymmetric Levi-Civita tensor,
\beq
n^{\mu} = \epsilon^{\mu \mud \dots \mu_n}\,\pdd \, \dots p_{n}\, .
\eeq
By construction $n^{\mu}$ is orthogonal to all of the other $n-1$ momenta.
In our case we have $n=d=4$, so we can define our vector $n^{\mu}$ in terms of $\pdd$, $\pt$ and $\pq$.

Our Gram matrix is given by 
\beq
\left(
\begin{array}{cccc}
\pdd^2 & \pdd \cdot \pt & \pdd \cdot \pq & 0 \\
\pt \cdot \pdd & \pt^{2} & \pt \cdot \pq & 0 \\
\pq \cdot \pdd & \pq \cdot \pt & \pq^{2} & 0 \\
0 & 0 & 0 & n^{2}
\end{array}
\right) \, , 
\eeq
with $n^2 = (\pdd\cdot \pt)^{2}\,\pq^{2} + (\pdd\cdot \pq)^{2}\,\pt^{2} + (\pt \cdot \pq)^{2}\,\pdd^{2} - 2\, \left(\pdd\cdot\pt\right)\, \left(\pdd\cdot\pq\right)\, \left(\pt\cdot\pq\right) - \pdd^2\,\pt^2\,\pq^{2}$.

If we compute the inverse Gram matrix and apply (\ref{metricton}) to $D_G^{\muu\nuu\mud\nud\mut\nut\muq\nuq}(\pu,\pdd,\pt,\pq)$, 
we can check that, indeed, the Euler counterterm vanishes. 

In our notebook, we first check that all of the traces of the re-expressed Euler counterterm are zero,
if up to two indices pair are left open, i.e.
\bea
\delta^{(4)}_{\muu\nuu} \, \delta^{(4)}_{\mud\nud} \,\delta^{(4)}_{\mut\nut} \, \delta^{(4)}_{\muq\nuq} \, D_{G}^{\muu\nuu\mud\nud\mut\nut\muq\nuq}(\pu,\pdd,\pt,\pq) &=& 0 \, , 
\nn \\
\delta^{(4)}_{\mud\nud} \,\delta^{(4)}_{\mut\nut} \, \delta^{(4)}_{\muq\nuq} \, D_{G}^{\muu\nuu\mud\nud\mut\nut\muq\nuq}(\pu,\pdd,\pt,\pq) &=& 0 \, , 
\nn \\
\delta^{(4)}_{\mut\nut} \, \delta^{(4)}_{\muq\nuq} \, D_{G}^{\muu\nuu\mud\nud\mut\nut\muq\nuq}(\pu,\pdd,\pt,\pq) &=& 0 \, .
\eea
These preliminary tests are not very informative, in fact, for if we consider the trace Ward identities for the Euler counterterm in $d$ dimensions, we see from (\ref{trace2tcount}),
(\ref{trace3tcount}) and (\ref{trace4tcount}) that the traces with four, three and two traced indices pairs must vanish also for $d\neq4$. On the other hand, if we have only one
traced pair, we see that the rhs of (\ref{trace4tcount}) in $d=4$ is given by a symmetric combination of the 3-point Euler counterterms.
Thus, proving that
\beq
\delta^{(4)}_{\muu\nuu} \, D_{G}^{\muu\nuu\mud\nud\mut\nut\muq\nuq}(\pu,\pdd,\pt,\pq) = 0 \, .
\label{euler3tnull}
\eeq
actually holds is the first non trivial test one can make.
One can observe that since the symmetric combination in the rhs of (\ref{trace4tcount}) certainly does not vanish
in $d \neq 4 $, then (\ref{euler3tnull}) is a non trivial check of the fact that also the $3$-point function counterterm must vanish for $d=4$. 
Indeed, its explicit form involves the projector onto the space of fully antisymmetric $5$-indices tensors~\cite{Osborn:1993cr}, which is necessarily zero for integer $d<5$.
This is also explicitly shown in our code.

Finally, we perform the same check for the fully uncontracted Euler counterterm, actually finding out that
\beq
D_{G}^{\muu\nuu\mud\nud\mut\nut\muq\nuq}(\pu,\pdd,\pt,\pq) = 0 \, .
\eeq
The full tests with all the necessary comments and details can be found in our repository.

\section{Functional derivatives}
\label{Vertices}

\subsection{Basic functional derivatives}

Here we give the basic momentum space functional derivatives, which are the building blocks needed to for all of the counterterms, vertices and anomalies. 
Once these few formulas are understood, the construction of all the rest is a matter of (very) careful bookkeeping
and a lot of applications of the chain rule for functional derivatives: please notice that in the following formulas
terms which vanish in the flat spacetime limit are already dropped, but one must take them into account for higher order derivatives.
\bea
\left[g_{\alpha\beta}\right]^{\muu\nuu} &=& \frac{1}{2}\, \left( \delta^{\muu}_\alpha\,\delta^{\nuu}_\beta + \delta^{\nuu}_\alpha\,\delta^{\muu}_\beta \right) \, ,
\nn \\
\left[g^{\alpha\beta}\right]^{\muu\nuu} &=& - \frac{1}{2} \left( \delta^{\muu\alpha} \,\delta^{\nuu\beta} + \delta^{\nuu\alpha}\,\delta^{\muu\beta} \right) \, ,
\nn \\
\left[\sqrt{g}\right]^{\muu\nuu} &=& \frac{1}{2}\, g^{\muu\nuu} \, ,
\nn \\
\left[\Gamma^\alpha_{\beta\chi}\right]^{\mu\nu}(k) 
&=& 
\frac{i}{2}\, \delta^{\alpha\lambda}\left( - \left[g_{\beta\chi} \right]^{\mu\nu} k_\lambda + \left[g_{\beta\lambda} \right]^{\mu\nu} k_\chi + \left[g_{\lambda\chi} \right]^{\mu\nu} k_\beta \right) \, , 
\nn \\
\left[\Gamma_{\alpha\beta\chi}\right]^{\mu\nu}(k) &\equiv& \delta_{\alpha\delta}\, \left[\Gamma^\delta_{\beta\chi}\right]^{\mu\nu}(k)
\nn \\
\left[\Gamma^\alpha_{\beta\chi}\right]^{\muu\nuu\mud\nud}(\ku,\kd)
&=& 
\left[g^{\alpha\lambda}\right]^{\muu\nuu}\, \left[\Gamma_{\lambda\beta\chi}\right]^{\mud\nud}(\kd) + (1\rightarrow 2) \, ,
\nn \\
\left[\Gamma^\alpha_{\beta\chi}\right]^{\muu\nuu\mud\nud\mut\nut}(\ku,\kd,\kt)
&=& 
\left[g^{\alpha\lambda}\right]^{\muu\nuu\mud\nud}\, \left[\Gamma_{\lambda\beta\chi}\right]^{\mut\nut}(\kt) + (1\leftrightarrow 2, 1\rightarrow 3) \, .
\eea
All further details are given in the file \emph{all\_functional\_derivatives\_computed.nb}

\subsection{Interaction vertices of the scalar field with gravitons}
%
We provide here the explicit forms of the three vertices used in the Feynman diagrams. 
The computation of the vertices can be done by taking at most three functional derivatives of the scalar action in curved space
with respect to the metric tensor and, of course, the scalar field.
This is because the vev's of the fourth order derivatives correspond to massless tadpoles, which are set to zero in DR,
so no fourth derivative is needed. 
\begin{itemize}
\item{graviton - scalar - scalar vertex}
\begin{figure}[h]
\begin{center}
\centerline{\includegraphics[scale=1.0]{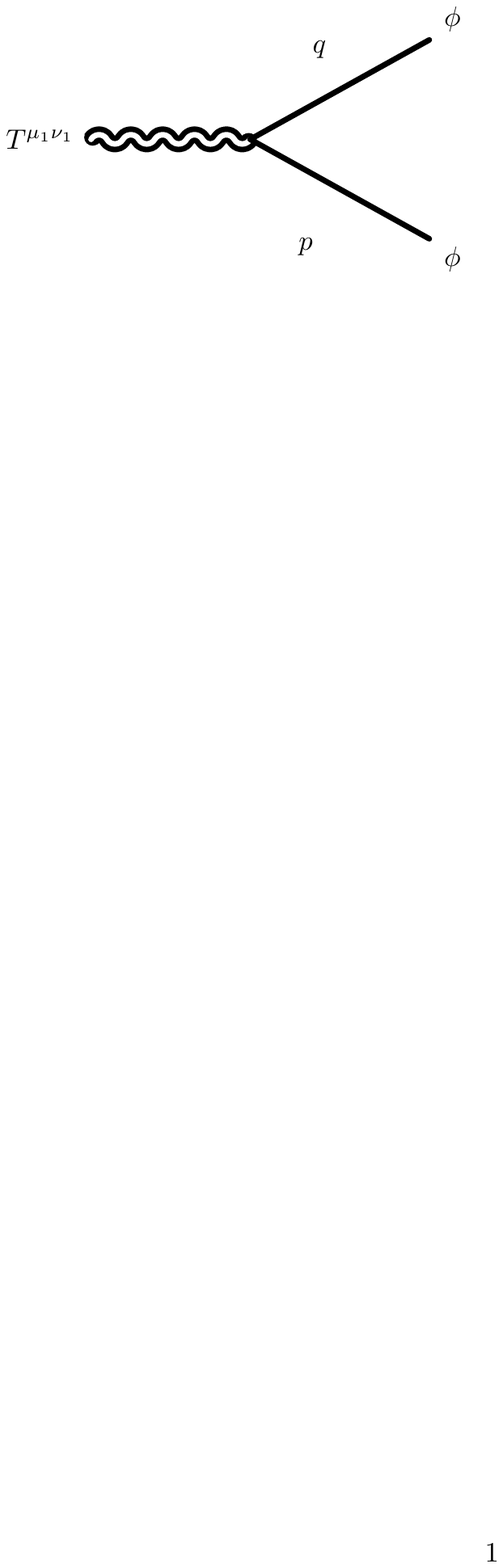}}
\end{center}
\end{figure}
\bea
V_{H\phi\phi}^{\mu\nu}(\vec{p},\vec{q}) 
&=&
\frac{1}{2}\left(\delta^{\mu\alpha}\delta^{\nu\beta} - \frac{1}{2}\delta^{\mu\nu}\delta^{\alpha\beta}\right)\, 
\left(p_{\alpha} q_{\beta} + p_{\beta} q_{\alpha}\right)\, ,
\nn \\
&+&  
\chi\, \left(\delta^{\mu\nu}\delta^{\alpha\beta} - \delta^{\mu\alpha}\delta^{\nu\beta}\right)\, 
\left(p_{\alpha}p_{\beta} + p_{\alpha}q_{\beta} + q_{\alpha}p_{\beta}+ q_{\alpha}q_{\beta} \right) \, , 
\nn
\eea
\item{graviton - graviton - scalar - scalar vertex}
\begin{figure}[h]
\begin{center}
\centerline{\includegraphics[scale=1.0]{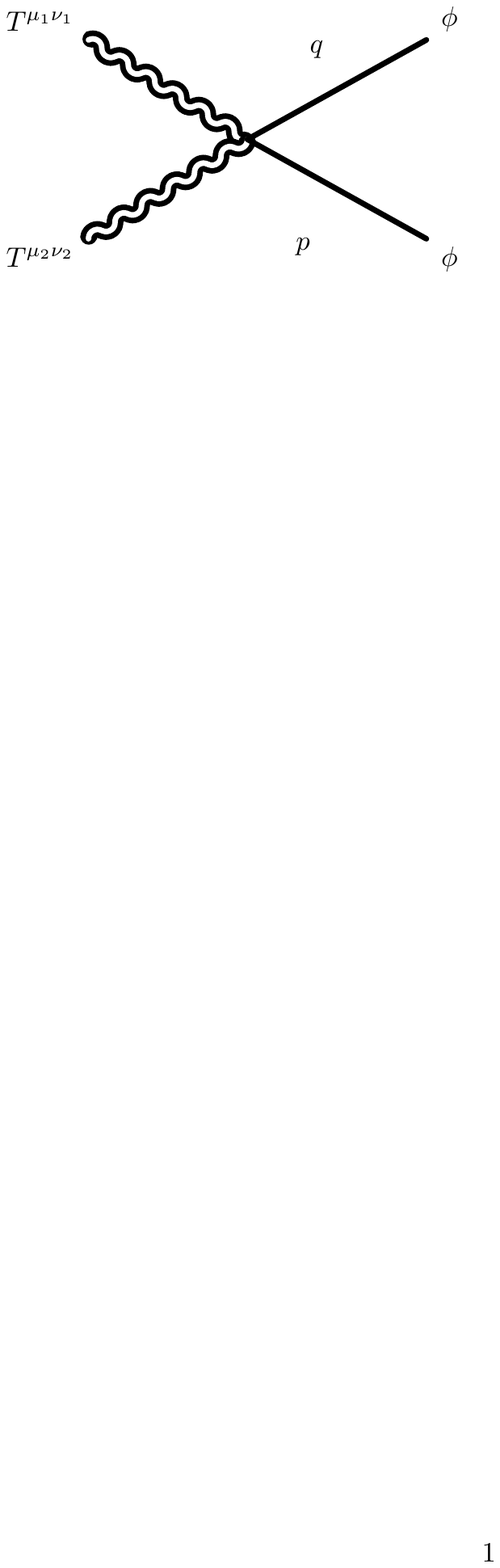}}
\end{center}
\end{figure}
\bea
V_{HH\phi\phi}^{\mu\nu\rho\sigma}(\vec{p},\vec{q},\vec{l}) 
&=&
\frac{1}{2}\, \bigg( \left[\sqrt{g}\right]^{\rho\sigma}\,
\left(\delta^{\mu\alpha}\delta^{\nu\beta} - \frac{1}{2}\delta^{\mu\nu}\delta^{\alpha\beta}\right)
\nn \\
&+& 
\left[g^{\mu\alpha}g^{\nu\beta} - \frac{1}{2}\, g^{\mu\nu}g^{\alpha\beta} \right]^{\rho\sigma}\,\bigg)
\left(p_{\alpha} q_{\beta} + p_{\beta} q_{\alpha}\right)\, ,
\nn \\
&+&
\chi\,\bigg\{ \bigg( \left[\sqrt{g}\right]^{\rho\sigma}\,
\left(\delta^{\mu\nu}\delta^{\alpha\beta} - \delta^{\mu\alpha}\delta^{\nu\beta}\right)
\nn \\
&+& 
\left[g^{\mu\nu}g^{\alpha\beta} - g^{\mu\alpha}g^{\nu\beta} \right]^{\rho\sigma}\,\bigg)
\left(p_{\alpha}p_{\beta} + p_{\alpha}q_{\beta} + p_{\beta}q_{\alpha} + q_{\alpha}q_{\beta}\right)
\nn \\
&+&
\left(\delta^{\mu\nu}\delta^{\alpha\beta} - \delta^{\mu\alpha}\delta^{\nu\beta}\right)\,
\left[\Gamma^\lambda_{\alpha\beta}\right]^{\rho\sigma}(\vec{l})\, i\, (p_\lambda + q_\lambda)
\nn \\
&-& 
\left(\frac{1}{2}\, \delta^{\mu\nu}\delta^{\alpha\beta} - \delta^{\mu\alpha}\delta^{\nu\beta}\right)\, 
\left[R_{\alpha\beta}\right]^{\rho\sigma}(\vec{l}) \bigg\} \, ,  
\nn
\eea
\item{graviton - graviton - scalar - scalar vertex}
\begin{figure}[h]
\begin{center}
\centerline{\includegraphics[scale=1.0]{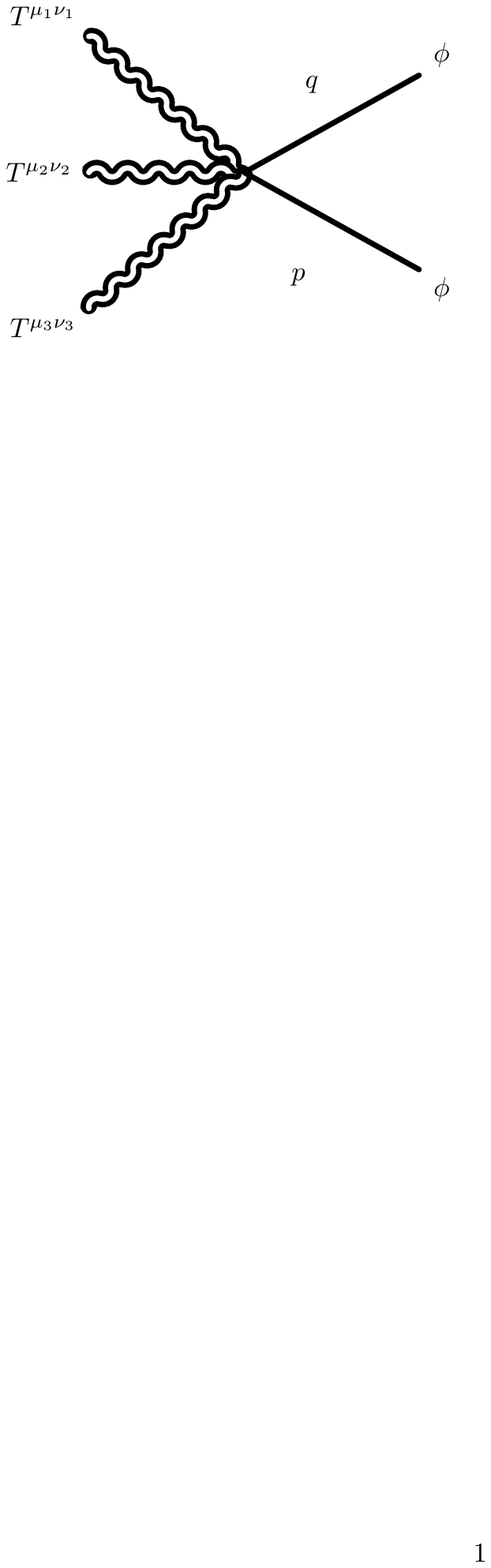}}
\end{center}
\end{figure}
\bea
V_{HHH\phi\phi}^{\mu\nu\rho\sigma\chi\omega}(\vec{p},\vec{q},\vec{l_1},\vec{l_2}) 
&=&
\frac{1}{2}\bigg\{\left[\sqrt{g}\right]^{\rho\sigma\chi\omega}\, 
\left(\delta^{\mu\alpha}\delta^{\nu\beta} - \frac{1}{2}\, \delta^{\mu\nu}\delta^{\alpha\beta}\right)
\nn \\
&+&
\left[\sqrt{g}\right]^{\rho\sigma}\, \left[g^{\mu\alpha}g^{\nu\beta} - \frac{1}{2}\, g^{\mu\nu} g^{\alpha\beta}\right]^{\chi\omega}+
\left[\sqrt{g}\right]^{\chi\omega}\, \left[g^{\mu\alpha}g^{\nu\beta} - \frac{1}{2}\, g^{\mu\nu} g^{\alpha\beta}\right]^{\rho\sigma} 
\nn \\
&+&
\left[g^{\mu\alpha}g^{\nu\beta} - \frac{1}{2}\, g^{\mu\nu} g^{\alpha\beta}\right]^{\rho\sigma\chi\omega} \bigg\}\, 
\left(p_{\alpha} q_{\beta} + p_{\beta} q_{\alpha}\right)\,
\nn \\
&+&
\chi\, \bigg\{\left[\sqrt{g}\right]^{\rho\sigma\chi\omega}\, 
\left(\delta^{\mu\nu}\delta^{\alpha\beta} - \delta^{\mu\alpha}\delta^{\nu\beta}\right)
\nn \\
&+&
\left[\sqrt{g}\right]^{\rho\sigma}\, \left[g^{\mu\nu}g^{\alpha\beta} - g^{\mu\alpha}g^{\nu\beta}\right]^{\chi\omega}	+
\left[\sqrt{g}\right]^{\chi\omega}\, \left[g^{\mu\nu}g^{\alpha\beta} - g^{\mu\alpha}g^{\nu\beta}\right]^{\rho\sigma} 
\nn \\
&+&
\left[g^{\mu\nu}g^{\alpha\beta} - g^{\mu\alpha}g^{\nu\beta}\right]^{\rho\sigma\chi\omega}
\bigg\}\, 
\left(p_\alpha p_\beta + p_\alpha q_\beta + q_\alpha p_\beta + q_\alpha q_\beta\right)
\nn \\
&+&
\chi\, \bigg\{
\left(\left[\sqrt{g}\right]^{\chi\omega}\, 
\left[\delta^{\mu\nu}\delta^{\alpha\beta} - \delta^{\mu\alpha}\delta^{\nu\beta}\right] 
+ \left[g^{\mu\nu}g^{\alpha\beta} - g^{\mu\alpha}g^{\nu\beta}\right]^{\chi\omega} \right)\, 
\left[\Gamma^\lambda_{\alpha\beta}\right]^{\rho\sigma}(\vec{l_1}) 
\nn \\
&+&
\left(\rho,\sigma,l_1\right) \leftrightarrow \left(\tau,\omega,l_2\right) 
+ \left(\delta^{\mu\nu}\delta^{\alpha\beta} - \delta^{\mu\alpha}\delta^{\nu\beta}\right)\, 
\left[\Gamma^\lambda_{\alpha\beta}\right]^{\rho\sigma\chi\omega}(\vec{l_1},\vec{l_2})
\bigg\}\, i\, \left(p_\lambda + q_\lambda\right)
\nn \\
&+&
\chi \bigg\{ 
\left(\left[\sqrt{g}\right]^{\chi\omega}\, \left( \delta^{\mu\alpha}\delta^{\nu\beta} -
\frac{1}{2}\, \delta^{\mu\nu}\delta^{\alpha\beta}\right)+
\left[g^{\mu\alpha}g^{\nu\beta} - \frac{1}{2}\,g^{\mu\nu}g^{\alpha\beta}\right]^{\chi\omega} \right)
\, \left[R_{\alpha\beta}\right]^{\rho\sigma}(\vec{l_1})
\nn \\
&+&
\left(\rho,\sigma,l_1\right) \leftrightarrow \left(\tau,\omega,l_2\right) 
+\left(\delta^{\mu\alpha}\delta^{\nu\beta} - \frac{1}{2}\,\delta^{\mu\nu}\delta^{\alpha\beta}\right) 
\left[R_{\alpha\beta}\right]^{\rho\sigma\chi\omega}(\vec{l_1},\vec{l_2}) \bigg\}
\nn
\eea
\end{itemize}
%

\providecommand{\href}[2]{#2}\begingroup\raggedright\endgroup

\end{document}